\documentclass[a4paper,11pt]{article}
\pdfoutput=1 

\usepackage{jcappub} 

\usepackage[T1]{fontenc} 
\usepackage{booktabs}
\usepackage{subcaption}
\usepackage{graphicx}
\usepackage{enumitem} 
\graphicspath{{./plots_paper/}}

\newcommand{\revise}{}

\title{Foreground removal with ILC methods for AliCPT-1}

\author[a,b]{Jiazheng Dou,}
\author[c]{Shamik Ghosh,}
\author[d]{Larissa Santos,}
\author[a,b]{Wen Zhao}
\affiliation[a]{CAS Key Laboratory for Researches in Galaxies and Cosmology, Department of Astronomy, University of Science and Technology of China, Chinese Academy of
Sciences, Hefei, Anhui 230026, P.R.China}
\affiliation[b]{School of Astronomy and Space Sciences, University of Science and Technology of China, Hefei 230026, P.R.China}
\affiliation[c]{Lawrence Berkeley National Laboratory,
Berkeley, CA 94720, U.S.A.}
\affiliation[d]
{Center for Gravitation and Cosmology, Yangzhou University, Yangzhou 224009, P.R.China}
\emailAdd{doujzh@mail.ustc.edu.cn, shamik-ghosh@outlook.com, larissa@yzu.edu.cn, wzhao7@ustc.edu.cn}

\abstract{One of the main goals of most future CMB experiments is the precise measurement of CMB B-mode polarization, whose major obstacle is the Galactic foregrounds. In this paper, we evaluate the foreground cleaning performance of the variants of the ILC method on partial sky B-modes and analyze the main sources of biases on the BB power spectrum. Specially, we compare the NILC, the cILC (in three domains) and the cMILC methods for AliCPT-1 simulations. We find that the cILC methods implemented in harmonic space and needlet space are both competent to clean different models of foregrounds, which bias the tensor-to-scalar ratio about 0.008 at maximum, and constrain the tensor-to-scalar ratio to $r<0.043\,$(95\%CL) for the AliCPT-1 configuration. 
We also note that the deviation of the estimated noise bias from the actual one for ILC, dubbed the noise bias error (NBE) in this paper, might make significant effects on the power spectrum for a small footprint and low signal-to-noise ratio CMB experiment. We finally obtain its relation with respect to the noise residual which fits well with the simulated results.}


\begin{document}
\maketitle
\flushbottom

\section{Introduction}
The Cosmic Microwave Background (CMB) anisotropies provide direct information about the origin and history of our universe. Observations of the CMB radiation in the last two decades (e.g., COBE \cite{bennettFourYearCOBEDMR1996}, WMAP \cite{bennettFirstYearWilkinsonMicrowave2003} and Planck \cite{aghanimPlanck2018Results2020a}) have been in excellent agreement with the $\Lambda$CDM model with very tight constraints on its parameters. However, despite $\Lambda$CDM success in explaining the observations, it has some well-known problems: horizon, flatness and magnetic monopole.  To alleviate them, the universe is supposed to undergo an inflationary epoch with a nearly exponential expansion. One of the most important predictions of cosmic inflation is the existence of primordial gravitational waves (PGW) generated from vacuum fluctuations in the early Universe (about $10^{-35}$s after the Big Bang). These PGWs would leave a faint imprint on the CMB polarization, known as primordial B-modes \cite{starobinskyPerturbationSpectrumEvolving1983}. The amplitude of the primordial B-mode power spectrum at degree angular scales is related to the amount of the PGWs, which is parameterized by the tensor-to-scalar ratio $r$. Thus, measuring $r$ would enable us to distinguish different classes of early Universe models (e.g.,  \cite{baumannCMBPolMissionConcept2009, martinEncyclopaediaInflationaris2014}). 

Attracted by the `smoking-gun' evidence of inflation, the measurement of the CMB primordial B-modes has been the main scientific goal of current CMB experiments, such as  POLARBEAR \cite{adachiMeasurementDegreeScale2020}, ACTpol \cite{louisAtacamaCosmologyTelescope2017}, SPTpol \cite{sayreMeasurementsBmodePolarization2020} and BICEP/Keck \cite{collaborationBICEPKeckXIII2021}. However, measuring the primordial CMB B-mode can be challenging due to the tiny signal and the various contaminants, such as the Galactic emissions, the instrumental noise and the CMB gravitational lensing, which converts E-modes into B-modes \cite{zaldarriagaGravitationalLensingEffect1998}. The current tightest upper limit of the tensor-to-scalar ratio, $r<0.032\,$(95\%CL), is given by a combination of Planck PR4 and BICEP2/Keck Array 2018 (BK18), BAO and CMB lensing data \cite{tristramImprovedLimitsTensortoscalar2022}.
Many on-going and future ground-based telescopes, as well as, space-based CMB experiments aim to reach higher sensitivity in order to be able to  measure the primordial B-modes, such as the Simons Observatory \cite{thesimonsobservatorycollaborationSimonsObservatoryScience2019}, QUBIC \cite{mennellaQUBICExploringPrimordial2019}, AliCPT \cite{liProbingPrimordialGravitational2019}, LiteBIRD \cite{hazumiLiteBIRDSatelliteStudies2019}, CMB-S4 \cite{collaborationCMBS4ForecastingConstraints2022}.

Contamination from polarized foreground signals, especially the Galactic thermal dust and the synchrotron emission, must then be disentangled from the B-mode signal by the so-called `component separation' methods. These methods are usually divided into blind, parametric and template removal techniques. Blind methods, such as ILC (Internal Linear Combination) \cite{eriksenForegroundRemovalWilkinson2004,tegmarkHighResolutionForeground2003}, FastICA (Fast Independent Component Analysis) \cite{mainoAllskyAstrophysicalComponent2002} and SMICA (Spectral Matching Independent Component Analysis) \cite{delabrouilleMultiDetectorMultiComponentSpectral2003,cardosoComponentSeparationFlexible2008}, usually rely on the statistical independence of components of various physical origins for their separation in multi-frequency observations, without assumptions of foregrounds. Parametric methods, such as Commander \cite{eriksenJointBayesianComponent2008} and FG-Buster \cite{stomporMaximumLikelihoodAlgorithm2009, errardFrameworkPerformanceForecasting2011}, model the foregrounds with spectral laws where parameters are fitted to the observed data. The parametric methods build an end-to-end Bayesian framework to estimate the parameters, but their effectiveness relies on the model. Template removal methods like SEVEM (Spectral Estimation Via Expectation Maximisation) \cite{fernandez-cobosMultiresolutionInternalTemplate2012} construct a set of foreground templates from the data.

The ILC method and its variants, such as the Needlet ILC (NILC) \cite{delabrouilleFullSkyLow2009}, play an important role in the CMB analysis of future experiments given our poor knowledge of the polarized foregrounds' properties. One of the caveats on the partial-sky analysis of B-modes is what is known as EB leakage \cite{zhaoSeparatingTypesPolarization2010,ghoshEndingPartialSky2021}: the decomposition from partial-sky QU maps into E- and B-modes will give rise to a considerable leakage from E-modes to B-modes. Directly applying ILC to Q and U maps separately is conceptually problematic since subtracting a constant contribution from Q or U is equivalent of introducing a pattern on the E and B modes. One should either linearly combine the QU maps to obtain the cleaned QU signals, e.g. PILC (Polarization ILC) \cite{fernandez-cobosExploringTwospinInternal2016}, or try to reconstruct the pure B-modes from QU maps before applying the ILC methodology. The PILC (constructing a complex-weight linear combination of $Q\pm iU$ such that the quantity $\langle P^2=Q^2+U^2\rangle$ of the final map is minimized) makes a tradeoff between minimizing the variance of E-modes and B-modes. Thereby, it may not be effective as a cleaning method to be applied in the CMB B-modes map recovery. Therefore, we adopt a novel template cleaning method \cite{liuMethodsPixelDomain2019} to extract the pure B-modes maps, and then apply several component separation methods on the multi-frequency B maps to clean the foregrounds.

Some authors already forecast on the performance of the standard ILC and NILC on B-modes mock data for some future experiments  \cite{remazeillesExploringCosmicOrigins2017,remazeillesPeelingForegroundsConstrained2021,adakBModeForecastCMBBh2021,krachmalnicoffInflightPolarizationAngle2022,zhangEfficientILCAnalysis2022,caronesAnalysisNILCPerformance2023, aurlienForegroundSeparationConstraints2023,fuskelandTensortoscalarRatioForecasts2023}. However, considering smaller sky fractions and lower sensitivities, such blind methods may give rise to a foreground-induced bias of the order $r\sim0.01$ \cite{remazeillesPeelingForegroundsConstrained2021,caronesAnalysisNILCPerformance2023}. Semi-blind extensions of ILC, such as cILC (constrained ILC) \cite{remazeillesCMBSZEffect2011} and cMILC (constrained Moment ILC) \cite{remazeillesPeelingForegroundsConstrained2021}, are proposed to project out some foreground components by constraining the multi-frequency weights using the assumed-known average of the spectral energy distribution (SED). This methodology is able to effectively reduce the foreground bias with the disadvantage of rising the uncertainty. In this paper, we consider both NILC and cILC algorithms implemented in the pixel, harmonic and needlet domains, as well as, the cMILC implemented in the needlet domain.

In our analysis, we use the mock data of seven frequency bands combining the WMAP K-band, the 4 Planck HFI bands and the 2 AliCPT-1 bands, which in total spans from 23GHz to 353GHz. The AliCPT (Ali CMB Polarization Telescope), a ground-based experiment located in Tibet, will measure CMB polarization in the northern hemisphere at 90GHz and 150GHz \cite{liProbingPrimordialGravitational2019}. The focus is on the trapezoid-shaped sky patch at the northern hemisphere with $f_{\rm sky}\sim7\%$, centering at ${\rm RA}=170^\circ$ and ${\rm DEC}=40^\circ$ for the Celestial coordinate.

Our paper is organized as follows. In Section~\ref{sec:methods}, we introduce the standard ILC method and its variants. In Section~\ref{sec:sims}, we describe the sky simulations used in our analysis. In Section~\ref{sec:pipeline}, we present the implementation details of the cILC method. In Section~\ref{sec:results}, we show the results of B-mode reconstruction and analyze the sources of biases. Finally, we conclude in Section~\ref{sec:conclusions}.

\section{Methods}
\label{sec:methods}
\subsection{Standard ILC}

The ILC method is widely used in recovering the CMB map from the observed sky using several frequency channels without any assumption of foreground emissions and noise. The cleaned CMB map is constructed as a linear combination of the multi-frequency data where the variance is minimized, leaving the CMB component unchanged.

First we model the observed sky as a linear mixture of different emissions. For the $c^{\rm th}$ emission component, let us assume it can be decomposed into a spatial template $s_c(p)$, which varies with the position in the sky, and a SED $a_c(\nu)$, which only depends on the frequency. The model for the observed data in the frequency band $\nu$, where $\nu\in\{1,2,\dots,n_\nu\}$, at pixel $p$ can be written as:
\begin{equation}
    \boldsymbol{d}(p) = \boldsymbol{A s}(p) + \boldsymbol{n}(p)\,,
    \label{eq:model}
\end{equation}
where $\boldsymbol{n}(p)=[n(\nu,p)]^T$ is a $n_\nu\times1$ column vector of multi-frequency instrumental noise, $\boldsymbol{s}(p)=[s_c(p)]^T$ is a $n_c\times1$ column vector of multi-component templates, and $\boldsymbol{d}(p)=[d(\nu, p)]^T$ is a $n_\nu\times1$ column vector of the observed data. $\boldsymbol{A}$ denotes the $n_\nu\times n_c$ mixing matrix of the emission components with elements $a_c(\nu)$. Note that hereafter, the observed data is already deconvolved to a common beam and resolution: $d_{\ell m}^\nu=\frac{b_\ell^0}{b_\ell^\nu}d_{\ell m}^{\nu,obs}$, where $b_\ell^{0}$ and $b_\ell^{\nu}$ are the common beam window function and the beam for each frequency band, respectively.  $d_{\ell m}^\nu$ is the harmonic coefficients of $d(\nu, p)$, and $d_{\ell m}^{\nu,obs}$ is the harmonic coefficients of the raw data.

The ILC method defines the estimated CMB map as a linear combination of all observed frequency maps with a constraint of weights $\sum_{\nu} w_\nu(p)=1$:
\begin{equation}
    \begin{aligned}
    \hat s_{\rm CMB} (p)&= \boldsymbol{w}^T(p)\cdot \boldsymbol{d}(p)  \\
    &= s_{\rm CMB} (p) + \sum_{\nu} w_\nu(p) \sum_{c} a_c(\nu)s_c(p) + \sum_{\nu} w_\nu(p) n(\nu, p)\,,
    \label{eq:ilc-cleaned-map}
    \end{aligned}
\end{equation}
where $\hat s_{\rm CMB} (p)$ is the estimate of the true CMB anisotropies $s_{\rm CMB} (p)$, $\boldsymbol{w}(p)=[w_\nu(p)]^T$ is a $n_\nu\times1$ column vector, and $c$ sums over foregrounds in the second term. Here, we assume that the CMB signal is frequency independent after calibration with respect to CMB, i.e. $a_{\rm CMB}(\nu)=1$. The weights $w_\nu(p)$ are obtained by minimizing the ensemble variance of the ILC-cleaned map $\langle\hat s^2_{\rm CMB} (p)\rangle$, using the Lagrange multiplier method, when the variance of the error $(\hat s_{\rm CMB}-s_{\rm CMB})$ is therefore minimized. In practice, the variance localized at pixel $p$ is estimated empirically by the averaged variance of the observed data over the vicinity of the $p$-th pixel. Defining the $n_\nu\times1$ CMB mixing vector as $\boldsymbol a=[1,\dots,1]^T$, the ILC weights are given as:
\begin{equation}
    \boldsymbol{w}^{\rm ILC}(p) = \frac{\boldsymbol {\hat C}^{-1} \boldsymbol a}{\boldsymbol a^T\boldsymbol {\hat C}^{-1} \boldsymbol a}\, ,\ \text{where}\ \boldsymbol{\hat C}(p)=\frac{1}{N_{p'}}\sum_{p'\in V(p)}q(p',p)\boldsymbol{d}(p')^\dagger \boldsymbol{d}(p')\,.
    \label{eq:pilc-wgts}
\end{equation}
Here $\boldsymbol {\hat C}$ is the $n_\nu\times n_\nu$ covariance matrix of the data as an estimate of the real covariance matrix, $V(p)$ is the vicinity of the pixel $p$, and $q(p',p)$ is some type of weight of $p'$ around the $p$-th pixel (for instance a Gaussian smooth kernel centered at $p$).

The method above implemented in pixel space is denoted as pILC. Likewise, the ILC implemented in harmonic space is denoted hILC, just replacing the pixel $p$ to the harmonic mode $(\ell,m)$, and the vicinity $V(p)$ to the binning of $(\ell,m)$s around $\ell$. The hILC weights are given by:
\begin{equation}
    \boldsymbol{w}_{\ell} = \frac{\boldsymbol {\hat C}_\ell^{-1} \boldsymbol a}{\boldsymbol a^T\boldsymbol {\hat C}_\ell^{-1} \boldsymbol a}\, ,\ \text{where}\ \boldsymbol{\hat C}_\ell=\sum_{\ell'\in B(\ell)}\sum_{m=-\ell'}^{\ell'}\boldsymbol{d}_{\ell'm}^\dagger \boldsymbol{d}_{\ell'm}\,.
    \label{eq:hilc-wgts}
\end{equation}
The covariance matrix $\boldsymbol{\hat C}_\ell$ is summed over the bin $B(\ell)$ centered at $\ell$ and the foreground-cleaned CMB map is $\hat s_{\ell m}=\boldsymbol{w}^T_{\ell}\cdot \boldsymbol{d}_{\ell m}$. The ILC in needlet domain is introduced in next section.

\subsection{NILC}
The needlet ILC, as a refinement of ILC, has been applied to extract CMB T-modes and E-modes for various experiments \cite{basakNeedletILCAnalysis2012,basakNeedletILCAnalysis2013,planckcollaborationPlanck2018Results2020a}.
The sky map of each channel is decomposed into a set of harmonic-filtered maps, which are localized in harmonic domain, and for each filtered map a perfect localization in pixel domain can be done by ILC independently. The observed data is filtered by the needlet bands:
\begin{equation}
    d^{\nu, j}_{\ell m} = h^j_\ell d^\nu_{\ell m}\,,
\end{equation}
where the needlet bands $h^j_\ell$ satisfying $\sum_j (h^j_\ell)^2=1$ are responsible for the localization in harmonic space. The cosine and Gaussian needlet bands are commonly used in NILC.

Assuming \texttt{HEALPix} pixelization \cite{gorskiHEALPixFrameworkHigh2005}, the spherical needlet function is defined as:
\begin{equation}
    \psi_{jk}(p) = \sqrt{\frac{4\pi}{N_j}}\sum_{\ell m} h^j_\ell Y_{\ell m}(p) Y^*_{\ell m}(n_{jk})\,,
\end{equation}
where $N_j$ is the \texttt{NPIX} parameter of the $j$-th needlet map, and $n_{jk}$ refers to the $k$-th pixel of the $j$-th needlet map. The $j$-th needlet map transformed from the observed data $d(\nu, p)$ is then given by:
\begin{equation}
    \begin{aligned}
    b^\nu_j(n_{jk}) &= \frac{4\pi}{N_p}\sum_{p} d(\nu, p) \psi^*_{jk}(p) \\
        &= \sqrt{\frac{4\pi}{N_j}}\sum_{\ell m} h^j_\ell d^\nu_{\ell m} Y_{\ell m}(n_{jk})\,,
    \end{aligned}
\end{equation}
while its inverse transformation is given by:
\begin{equation}
    d^\nu_{\ell m} = \sum_{jk}b^{\nu}_j(n_{jk}) \sqrt{\frac{4\pi}{N_j}} h^j_\ell Y^*_{\ell m}(n_{jk})\,.
\end{equation}

The needlet transformation is a linear operation. The final NILC map is transformed from the individually-cleaned needlet maps, which are linearly combined by the input multi-frequency needlet maps for each needlet band $j$:
\begin{subequations}
    \begin{align}
    &\hat s^{\rm NILC}_{\ell m} = \sum_{jk}b^{\rm NILC}_j(n_{jk}) \sqrt{\frac{4\pi}{N_j}} h^j_\ell Y^*_{\ell m}(n_{jk})\, ,
    \\
    &b^{\rm NILC}_j(n_{jk}) = \sum_{\nu} w^{\rm NILC}_{\nu, j}(n_{jk}) b^\nu_j(n_{jk})\,.
    \end{align}
\end{subequations}
Just as the pixel domain ILC method, the NILC weights are given by:
\begin{equation}
    \boldsymbol{w}^{\rm NILC}_{j}(n_{jk}) = \frac{\boldsymbol {\hat C}_{jk}^{-1} \boldsymbol a}{\boldsymbol a^T\boldsymbol {\hat C}_{jk}^{-1} \boldsymbol a}\,.
    \label{eq:nilc-wgts}
\end{equation}

To estimate the covariance matrices for the $k$-th pixel of the $j$-th needlet scale $\boldsymbol C_{jk} = C^{\nu_1\times \nu_2}_{jk} = \langle b^{\nu_1}_j(n_{jk}) b^{\nu_2}_j(n_{jk}) \rangle$, we compute the empirical covariance $\boldsymbol{\hat C}_{jk}$ by averaging the needlet coefficient products $b^\nu_j(n_{jk}) b^\nu_j(n_{jk})$ over some domain of pixels around pixel $k$, which can be written as:
\begin{equation}\label{eq:nilc-cov}
    \hat C^{\nu_1\times \nu_2}_{jk} = \frac{1}{N_k} \sum_{k'} q_j(k, k') b^{\nu_1}_j(n_{jk'}) b^{\nu_2}_j(n_{jk'})\,,
\end{equation}
where $q_j(k, k')$ are weights dependent on the needlet scale. More details are introduced in Section~\ref{sec:cNILC-imp}.

The final NILC cleaned map is inverse spherical harmonic transformed (SHT) to pixel space as: 
\begin{equation}
    \hat s_{\rm NILC} (p)= \sum_{\ell m} \hat s^{\rm NILC}_{\ell m} Y_{\ell m} (p) \,.
\end{equation}

\subsection{Constrained ILC}
The constrained ILC (cILC) method \cite{remazeillesCMBSZEffect2011} introduces an assumption about the SED of the main foreground components. By construction, it adds constraints to cancel those unwanted components.  However, while this process reduces the residual foregrounds, it increases the noise level as an expense.

Given the mixing matrix, the constraints are given by:
\begin{equation}
    \sum_\nu w^{\rm cILC}_\nu a_c(\nu) = e_c\,,
\end{equation}
where $e_c$ is $1$ for CMB and $0$ for all foreground components modeled in the mixing matrix. The cILC weights are given by:
\begin{equation}\label{eq:cilc-weights}
    \boldsymbol{w}^{T} = \boldsymbol{e}^T \left( \boldsymbol{A}^T\hat{\boldsymbol{C}}^{-1} \boldsymbol{A}\right)^{-1}\boldsymbol{A}^T \hat{\boldsymbol{C}}^{-1}\,,
\end{equation}
where $\hat{\boldsymbol{C}}$ is the covariance matrix of the data. The cILC cleaned CMB map is given by $\hat s_{\rm CMB}=\boldsymbol{w}^T\cdot \boldsymbol{d}$.

As the ILC, the cILC can also be implemented in pixel, harmonic and needlet domain, hereafter denoted by cpILC, chILC and cNILC, respectively.

\subsubsection{The mixing matrix}
The mixing matrix used in cILC is determined by the assumed SEDs of the foreground components of interest, which, in our analysis, are synchrotron and thermal dust. We model the SED of CMB anisotropies as a differential black body, of synchrotron emission as a power law with a fixed spectral index of $\beta_{\rm sync}=-3$, and of the thermal dust as a modified black body with the dust temperature of $\rm T_{dust}=19.6K$ and the dust spectral index of $\beta_{\rm dust}=1.59$, for data in Rayleigh-Jeans brightness temperature units. These SED parameters are from the Planck constraints on polarized foregrounds \cite{planckcollaborationPlanck2018Results2020e,planckcollaborationPlanckIntermediateResults2016a}. The SEDs after calibration to the CMB thermodynamic temperature are written as:
\begin{equation}
    \left\{
        \begin{aligned}
            & a_{\rm CMB}(\nu)=1\,,\\
            & a_{\rm sync}(\nu)=\frac{g_\nu}{g_{\nu_s}}(\frac{\nu}{\nu_s})^{\beta_s} \,,\\
            & a_{\rm dust}(\nu)=\frac{g_\nu}{g_{\nu_d}}(\frac{\nu}{\nu_d})^{\beta_d+1}\frac{\exp{(x_d(\nu_d))}-1}{\exp{(x_d(\nu))}-1}\,,
        \end{aligned}
    \right.
\end{equation}
where $x_d(\nu)\equiv\frac{h\nu}{k_B T_d}$, $\nu_s=23$GHz and $\nu_d=353$GHz are the reference frequencies of synchrotron and dust emissions, respectively. $g_\nu$ is the unit conversion factor from $\mu K_{\rm RJ}$ to $\mu K_{\rm CMB}$.

In principle, the adopted SED parameters should fit the real sky foregrounds to optimize the foreground cleaning, and should be position-dependent in the best case for the actual data with parameters varying over the sky. On the other hand, given lack of prior knowledge of the polarized foregrounds, the mixing matrix should be robust enough to be applied on different sky models. Therefore, in Section~\ref{sec:res-fg} we test the cILC method on four different foreground models (introduced in Section~\ref{sec:fg}).

\subsection{Extensions: constrained Moment-ILC}
\label{sec:cMILC}
The constrained Moment-ILC (cMILC) \cite{remazeillesPeelingForegroundsConstrained2021} is an extension of cILC that adds several nulling constraints on the SEDs of the main moments of the foreground emissions. Assuming that the SED of the $c^{\rm th}$ component varies over the sky, we can expand the mixing amplitude around some fixed pivot SED parameters $\boldsymbol{\bar\beta}=(\bar\beta_1,\dots\bar\beta_n)^T$:
\begin{equation}
    a_c(\nu,\boldsymbol{\beta(p)})=\sum_k\sum_{\alpha_1+\cdots+\alpha_n=k}\frac{(\beta_1(p)-\bar\beta_1)^{\alpha_1}\cdots(\beta_n(p)-\bar\beta_n)^{\alpha_n}}{\alpha_1!\cdots\alpha_n!}\frac{\partial^k a_c(\nu,\boldsymbol{\bar\beta})}{\partial\bar\beta_1^{\alpha_1}\cdots\partial\bar\beta_n^{\alpha_n}}\,.
\end{equation}
Under the approximation of the finite-order expansion, the derivative terms can all be analytically computed in terms of the channel $\nu$ and the fixed $\boldsymbol{\bar\beta}$. Similar with the cILC method, we add constraints to null these high-order moments of the foreground SED:
\begin{equation}
    \sum_\nu w_\nu^{\rm cMILC} \frac{\partial^k a_c(\nu,\boldsymbol{\bar\beta})}{\partial\bar\beta_1^{\alpha_1}\cdots\partial\bar\beta_n^{\alpha_n}}=e_c\,.
\end{equation}
It can be seen that the constraint turns to that of cILC when $k=0$. The additional constraints (of $k=1,2,\dots$) further reduce the residual foreground with an expense of increasing the residual noise power. Besides, the number of observation channels cannot be less than the number of constraints to ensure the solvability of the weights.
In this work, with seven bands involved, cMILC is considered up to the first order ($k=1$) and implemented in needlet space with an additional constraint corresponding to the first order derivative of the thermal dust with respect to the dust temperature $T_d$. More specifically, the new constraint is given by:
\begin{equation}
    \sum_\nu w_\nu^{\rm cMILC} \frac{\partial a_{\rm dust}(\nu)}{\partial T_d}|_{\bar T_d}=0\,,
\end{equation}
where the pivot SED parameters are the same as the parameters of the mixing matrix.



\section{Simulations}
\label{sec:sims}
The simulated data sets are based on the first season observation of the future CMB experiment AliCPT, with Planck HFI four bands and WMAP K band combined as ancillary simulations to improve the efficiency of foreground removal. Unlike simulations in \cite{ghoshPerformanceForecastsPrimordial2022}, our maps are not generated from time-ordered data (TOD), and instrumental systematics like filtering effects are not involved in our analysis, which can be generally modeled as a set of transfer functions independent of the foreground removal process. Each realization includes observations of seven frequency channels: the 95GHz and 150GHz bands of AliCPT mission, four bands of Planck HFI instruments(100, 143, 217 and 353GHz), and the WMAP K band (23GHz). The input maps are the sum of CMB, noise and foregrounds smoothed with the instrumental Guassian beams information listed in Table~\ref{tab:instr}. We adopt a \texttt{HEALPix} $N_{side}=1024$ pixelization scheme.

\begin{table}[tbp]
    \centering
    \caption{Summary Table~of seven channels used in our simulations, including WMAP K-band, AliCPT 95GHz and 150GHz bands, Planck HFI bands. $\sigma_n^{\rm P}$ is the map noise level for polarization. Note that the noise levels are computed by fitting the noise power spectra in the sky patch with the white noise spectra, where the Planck levels are slightly lower than the values from Table 4 of \cite{planckcollaborationPlanck2018Results2020c}. The ideal channel-combined noise level is about 10$\mu$K-arcmin.}
    \begin{tabular}{lc|cc|cccc}
        \bottomrule
        Instruments&WMAP&\multicolumn{2}{c|}{AliCPT}&\multicolumn{4}{c}{Planck HFI} \\
        \hline
        Frequency(GHz)&23&95&150&100&143&217&353 \\
        \hline
        Beam size(arcmin)&52.8&19&11&9.7&7.3&5.0&4.9 \\
        \hline
        $\sigma_n^{\rm P}$($\mu$K-arcmin)&496&13&18&78&65&91&404\\
        \toprule
    \end{tabular}
    \label{tab:instr}
\end{table}

\subsection{Masks}
\begin{figure}[tbp]
    \centering
    \begin{subfigure}[b]{0.4\textwidth}
        \centering
        \includegraphics[width=\textwidth]{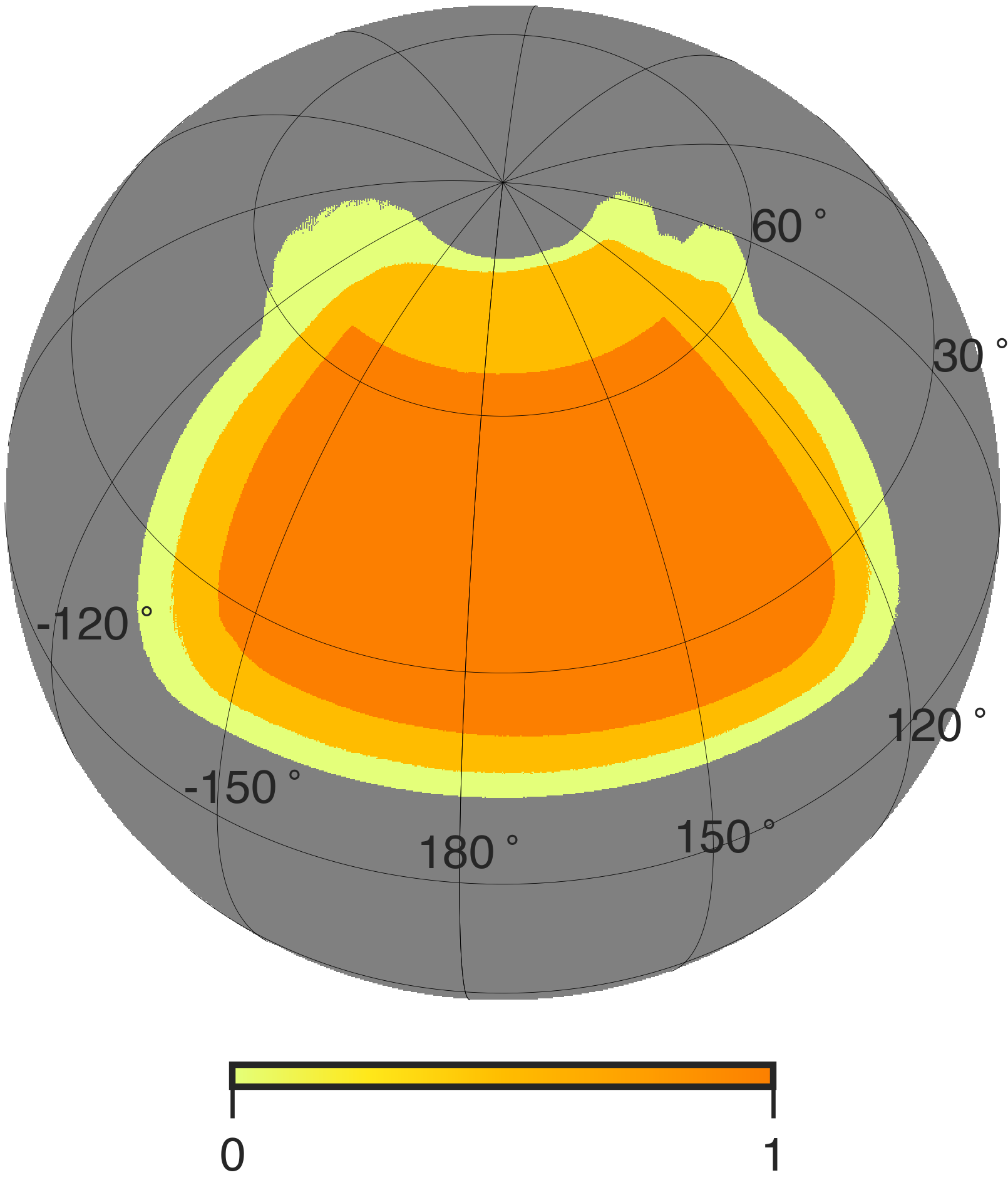}
        \caption{binary masks}
    \end{subfigure}
    \begin{subfigure}[b]{0.4\textwidth}
        \centering
        \includegraphics[width=\textwidth]{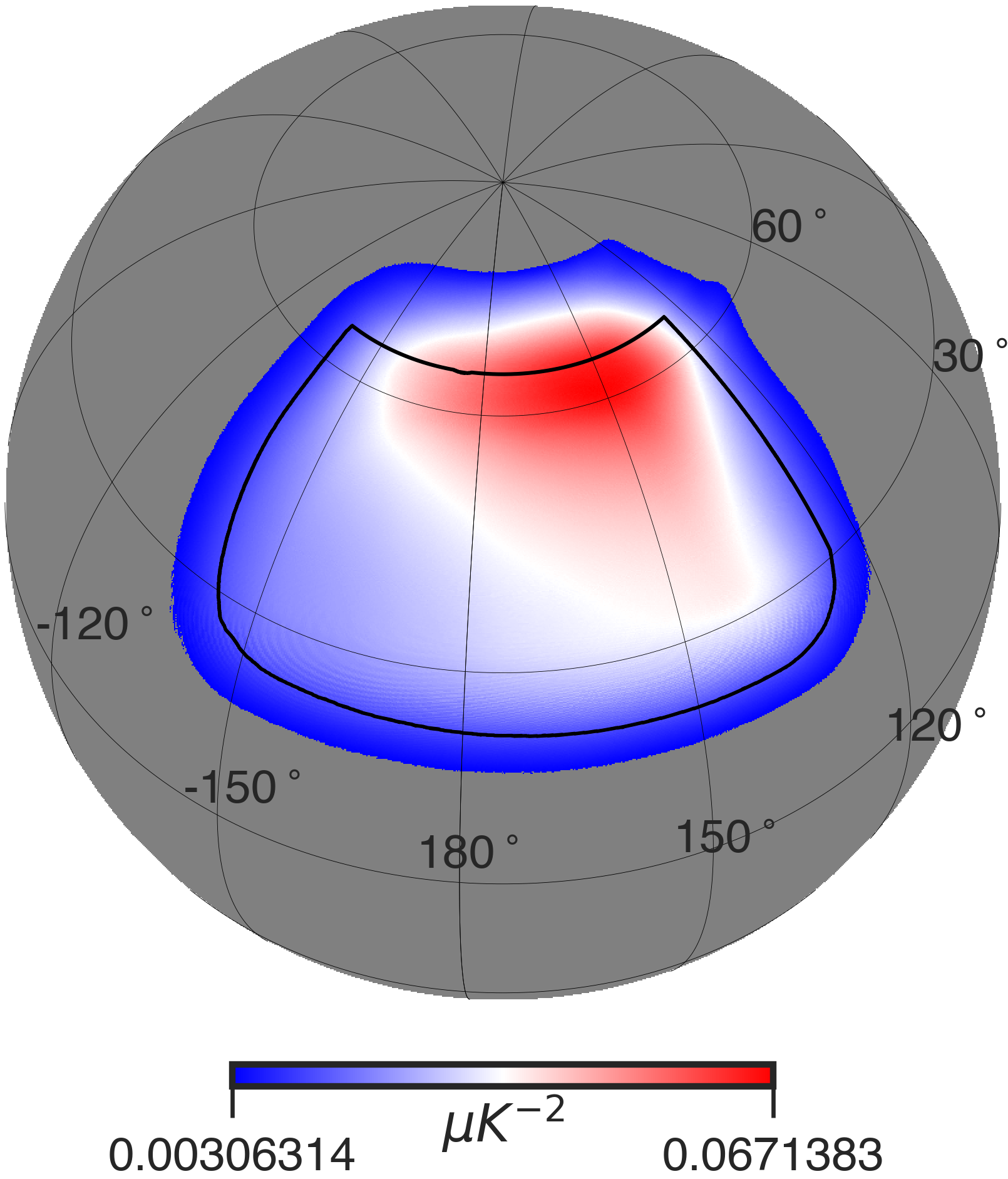}
        \caption{inverse variance}
    \end{subfigure}
    \caption{Masks in the AliCPT-1 data analysis. (a) From the outer to inner region, the figure shows the AliCPT-1 observation patch (yellow), the $20\mu$K mask (light orange), and the UNP mask( dark orange), with sky fractions of 13\%, 10\%, 6.7\%, respectively. (b)The inverse variance of the polarization noise at 150GHz. The black curve shows the boundary of the UNP mask, inside which is the UNP-inv mask.}
    \label{fig:msk}
\end{figure}

The AliCPT scanning region in the first observation season centers at ${\rm RA}=180^\circ$, ${\rm DEC}=30^\circ$ in the northern sky, covering about 17\% of the sky area, as shown in Figure~\ref{fig:msk}. In the foreground cleaning pipeline, we adopt two binary masks: the `$20\mu$K' mask and the `UNP' mask. They are shown in Figure~\ref{fig:msk}. The $20\mu$K mask is defined by masking the pixels with the noise standard deviation above $20\mu$K at \texttt{NSIDE}=1024 for the 150GHz channel. We then further remove the sky above declination of $65^\circ$, obtaining a mask with $f_{\rm sky}=6.7\%$ named the UNP mask. 
We also divide the UNP mask by the variance of the polarization noise at 150GHz, hereafter named the UNP-inv mask.

\subsection{CMB}
We produce two sets of lensed CMB data ($A_L=1$) with a fiducial tensor-scalar ratio of $r=0.03$ and the null value $r=0.0$, with each of the sets containing 300 seven-channel simulations. The CMB maps are Gaussian realizations generated from the power spectra generated by the \texttt{CAMB} \cite{lewisEfficientComputationCMB2000} package using the best-fit Planck 2018 parameters\footnote{Specifically, the parameters include dark matter dencity $\Omega_ch^2=0.120$, baryon density $\Omega_bh^2=0.02237$, scalar spectral index $n_s=0.9649$, optical depth $\tau=0.0544$, Hubble constant $H_0=69.36\ {\rm km\ s^{-1}\ Mpc^{-1}}$ and the primordial comoving curvature power spectrum amplitude $A_s=2.10\times10^{-9}$.} \cite{aghanimPlanck2018Results2020a}. The lensing effects are added by the \texttt{LensPyx} \cite{reineckeImprovedCMBLensing2023} package. Unless specified, the CMB realizations with $r=0.03$ are used in the following analysis.

\subsection{Noise}
\label{sec:noise}
The instrumental noises for AliCPT-1 and WMAP-K channels are Gaussian white realizations generated from the noise variance maps of AliCPT-1 bands and WMAP K band (upgraded to $N_{\rm side}=1024$), respectively. For Planck HFI four bands, we use Planck FFP10 noise simulations from the Planck Legacy Archive\footnote{http://pla.esac.esa.int/pla}.

\begin{figure}[tbp]
    \centering
    \begin{subfigure}[b]{0.7\textwidth}
        \centering
        \includegraphics[width=\textwidth]{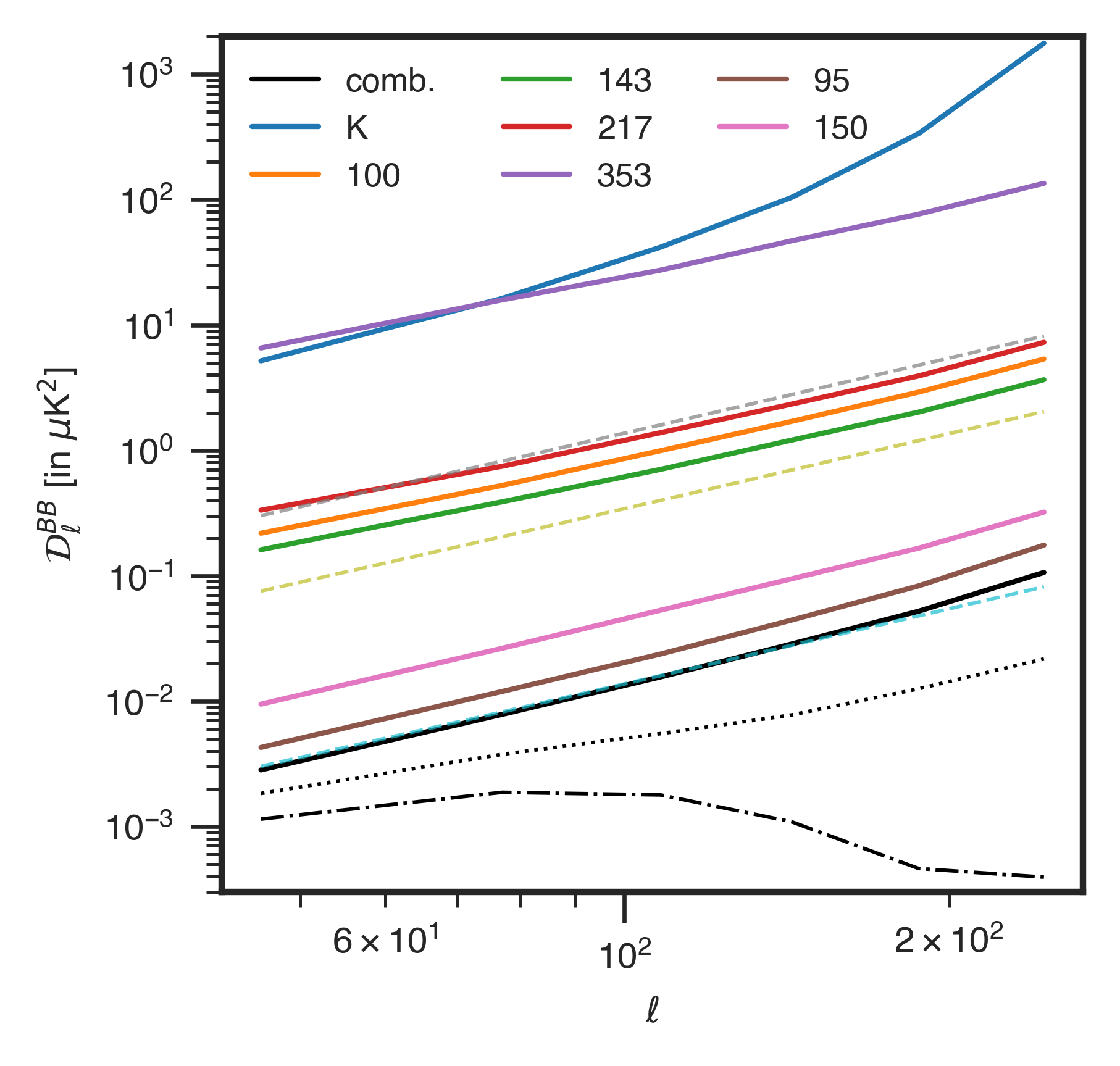}
    \end{subfigure}
    \caption{The debeamed polarization noise power spectra $\mathcal{N}_\ell/b_\ell^2$ of our seven frequency channels. For reference, the white noise spectra with noise levels of 10, 50 and 100 $\mu$K-arcmin are plotted as three dashed curves from bottom to top. The black solid curve is the channel-combined noise spectrum, which is about 10 $\mu$K-arcmin. The lensed CMB BB spectrum (black dotted curve) and the tensor BB spectrum with $r=0.03$ (black dot-dashed curve) are also plotted for comparison.}
    \label{fig:Nl}
\end{figure}

The debeamed polarization noise power spectra $\mathcal{N}_\ell/b_\ell^2=\frac{\ell(\ell+1)}{2\pi}{N}_\ell/b_\ell^2$ for seven channels are shown in Figure~\ref{fig:Nl}. We hereafter omit the $b_\ell^2$ factor since the noises are all debeamed in the final cleaned map. The ideal noise level combining the data of all channels is about 10 $\mu$K-arcmin, given by $\mathcal{N}_\ell=[\sum_\nu \mathcal{N}_{\ell,\nu}^{-1}]^{-1}$. 

\subsection{Foregrounds}
\label{sec:fg}
The diffuse foregrounds include various components, such as synchrotron, thermal dust, anomalous microwave emission (AME), free-free and CO emissions. However, the synchrotron emission and the thermal dust are known to be two main polarization contaminants among them. We do not consider the polarized point sources in our main analysis since they are not modeled in the cILC method, but we show in Appendix \ref{sec:ps} that their effects are negligible to the extent of current sensitivity. In order to properly assess the performance of cILC methods on various sky models, we use four Galactic models based on the Planck Sky Model (PSM) \cite{delabrouillePrelaunchPlanckSky2013} to generate our foreground simulations using the Python Sky Model (\texttt{pysm} \cite{zoncaPythonSkyModel2021}) package. The models are specified as follows:
\begin{itemize}[leftmargin=*]

\item[-] \emph{AliCPT foregrounds:} In this model, the thermal dust polarization maps are generated from the GNILC template of the Planck 2018 release \cite{planckcollaborationPlanck2018Results2020a}, scaled to different frequencies by a modified blackbody SED with the dust temperature and spectral indices from the GNILC Planck 2015 dust maps best-fit \cite{planckcollaborationPlanckIntermediateResults2016}. The synchrotron polarization template is based on the Planck 2018 SMICA map, and it is scaled by a power law SED with a fixed $\beta_s$ of -3.08. In this work, this is our baseline model.

\item[-] \emph{d0s0:} The \texttt{d0s0} \texttt{pysm} model uses the 353GHz map from Planck 2015 release as the thermal dust polarization template, and the WMAP 9-year 23 GHz Q/U map \cite{bennettNineYearWilkinsonMicrowave2013} as the synchrotron template, with fixed spectral indicies of $\beta_s=-3$, $\beta_d=1.54$ and $T_d=20K$. The polarization templates are smoothed with a Gaussian kernel of FWHM $2.6^\circ$ for thermal dust and $5^\circ$ for synchrotron, and have small scales added via the procedure described in \cite{thornePythonSkyModel2017}.

\item[-] \emph{d1s1:} The templates for the \texttt{d1s1} \texttt{pysm} model are the same as \texttt{d0s0}. The spatially varying spectral indices for thermal dust are obtained from the Planck 2015 Commander dust map \cite{planckcollaborationPlanck2015Results2016b}; for synchrotron $\beta_s$ is derived using a combination of the Haslam 408 MHz data and WMAP 23 GHz 7-year data \cite{miville-deschenesSeparationAnomalousSynchrotron2008}.

\item[-] \emph{d10s5:} The thermal dust maps are based on the Planck 2018 GNILC maps \cite{planckcollaborationPlanck2018Results2020a}, which present a larger variation of $\beta_d$ and $T_d$ parameters at low resolution than the \texttt{d1} model. The synchrotron templates are the same as \texttt{s1}, with the spectral index map from \texttt{s1} rescaled based on the S-PASS data \cite{krachmalnicoffSPASSViewPolarized2018}. Small-scale structures are added as Gaussian realizations of power-law power spectra. 
\end{itemize}

\begin{figure}[tbp]
    \centering
    \begin{subfigure}[b]{0.32\textwidth}
        \centering
        \includegraphics[width=\textwidth]{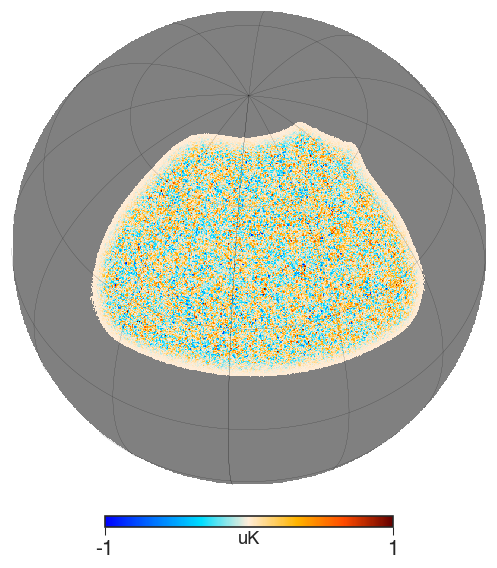}
        \caption{CMB}
    \end{subfigure}
    \begin{subfigure}[b]{0.32\textwidth}
        \centering
        \includegraphics[width=\textwidth]{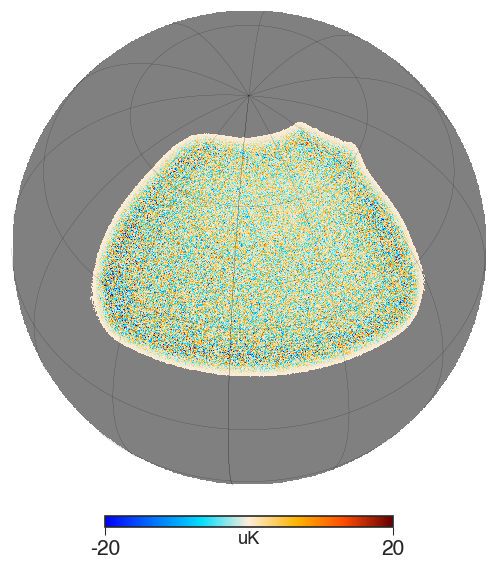}
        \caption{Noise}
    \end{subfigure}
    \begin{subfigure}[b]{0.32\textwidth}
        \centering
        \includegraphics[width=\textwidth]{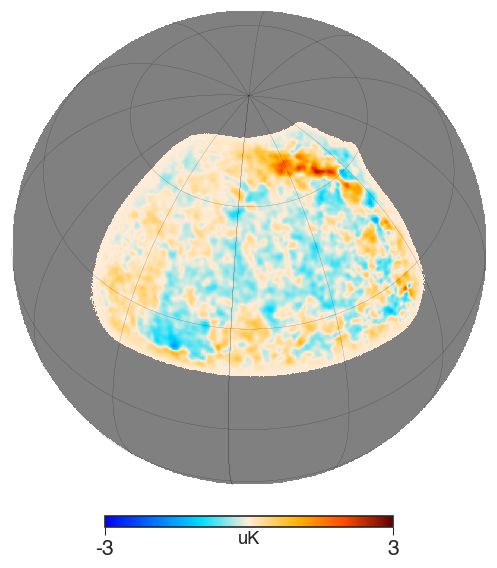}
        \caption{AliCPT FG}
    \end{subfigure}
    \\
    \begin{subfigure}[b]{0.32\textwidth}
        \centering
        \includegraphics[width=\textwidth]{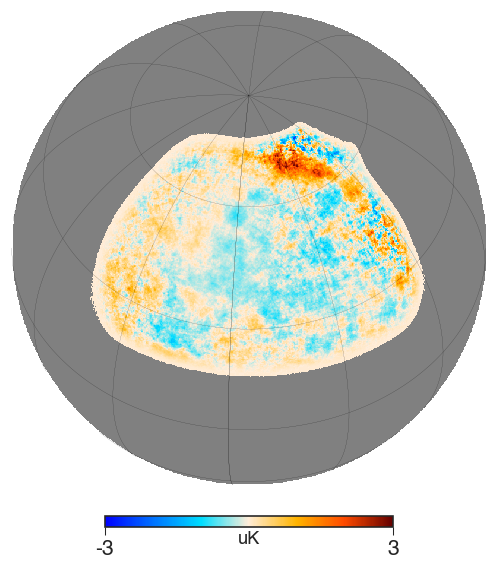}
        \caption{d0s0 FG}
    \end{subfigure}
    \begin{subfigure}[b]{0.32\textwidth}
        \centering
        \includegraphics[width=\textwidth]{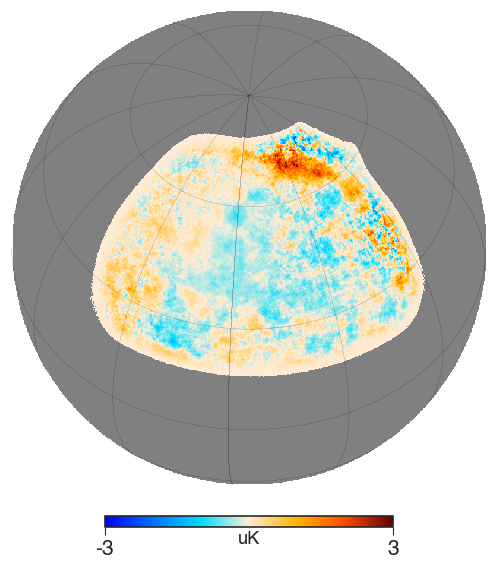}
        \caption{d1s1 FG}
    \end{subfigure}
    \begin{subfigure}[b]{0.32\textwidth}
        \centering
        \includegraphics[width=\textwidth]{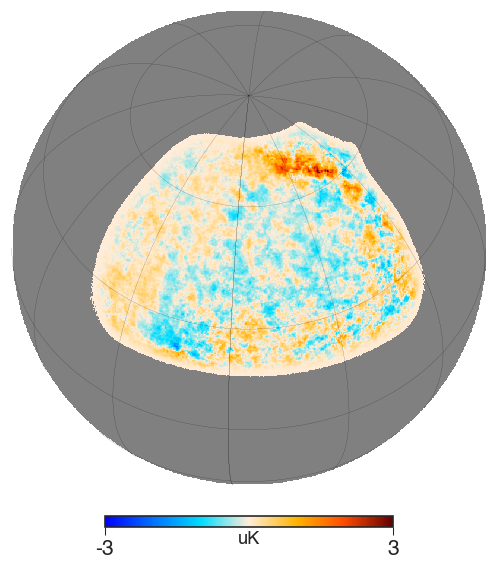}
        \caption{d10s5 FG}
    \end{subfigure}
    \caption{One sample of the simulated CMB, noise and foreground B-mode maps at 150GHz band after template cleaning with 6$^\circ$ C2 apodization of the $20\mu$K mask.}
    \label{fig:sample}
\end{figure}

Four sets of multi-frequency foregrounds are generated using the above models. A set of B-mode maps (CMB, noise and foreground) at 150GHz for AliCPT-1 is shown in Figure~\ref{fig:sample}. The B-mode maps are obtained by the template cleaning method (see Section~\ref{sec:pcl-tc}).

\section{Pipelines: CMB B-mode reconstruction}
\label{sec:pipeline}
We now implement the NILC, chILC, cpILC, cNILC and cMILC methods on the multi-frequency data sets, where the cMILC is implemented in needlet space. First we have to extract the pure B-mode maps from the input partial-sky QU maps considering the partial-sky effects. Then we apply the ILC cleaning process on the B-mode maps in pixel, harmonic or needlet domain. Finally, we compute the noise debiased power spectra and extract the tensor-to-scalar ratio value from it.

\subsection{EB leakage and Power spectrum estimation}
\label{sec:pcl-tc}
Two effects must be considered when reconstructing the full-sky power spectrum from a partial-sky map: the EB leakage, and the mode coupling of the pseudo power spectrum. The template cleaning method introduced by Liu et al. \cite{liuMethodsPixelDomain2019} is an efficient way to remove the EB leakage in pixel domain, which can suppress the leakage to a level of $r\sim 10^{-4}-10^{-5}$ over $\ell\sim60$.

The template cleaning steps are as follows. 
\begin{enumerate}
    \item The masked polarization map $\boldsymbol{P}=(Q,U)$ is decomposed into the so-called E and B family maps: $\boldsymbol{P_E}=(Q_E,U_E)$ and $\boldsymbol{P_B}=(Q_B,U_B)$, which is done by $(Q,U)\rightarrow E;B\rightarrow (Q_E,U_E);(Q_B,U_B)$.
    \item Obtain $\boldsymbol{P'_B}$ in the same way from the masked E family $\boldsymbol{P_E}$, i.e. $\boldsymbol{P_E}\rightarrow \boldsymbol{P'_E};\boldsymbol{P'_B}$.
    \item Use the masked $\boldsymbol{P'_B}$ as a E-B leakage template to remove the leakage from the masked $\boldsymbol{P_B}$ by linear fitting.
\end{enumerate}

After correcting the EB leakage and foreground cleaning, we compute the B-mode power spectrum with the pseduo-$C_\ell$ (PCL) estimator using the \texttt{NaMASTER} \cite{alonsoUnifiedPseudoC_2019}\footnote{https://github.com/LSSTDESC/NaMaster} python package, which takes the masking, binning, pixel-window and beaming effects into account.

In order to assess the accuracy of power spectrum estimation on foreground-free maps, we apply the template cleaning and the PCL process on 100 CMB-only simulations smoothed with Gaussian beams of either FWHM=52.8 arcmin or 11 arcmin as the possible common beams, corresponding to the largest beam among our frequency channels and the beam used in AliCPT for NILC \cite{hanForecastsCMBLensing2023}, respectively. First, we obtain the real B-mode maps from the full-sky CMB simulations as reference. The simulations using the $20\mu$K mask go through the template cleaning pipeline to correct for the leakage. The corrected B-mode maps are then masked again by the $20\mu$K mask with 6 degree C2 type apodization in order to reduce the high leakage residual on the boundary of the mask \cite{liuMethodsPixelDomain2019}. Finally, we apply PCL on the leakage-cleaned maps to reconstruct the full-sky power spectra. The estimated BB power spectra are binned with $\Delta\ell=\max\{30,0.3\ell_{\min}\}$ (where $\ell_{\min}$ is the starting multipole of a bin) from $\ell=40$ to 600.

\begin{figure}[tbp]
    \centering
    \begin{subfigure}[b]{0.49\textwidth}
        \centering
        \includegraphics[width=\textwidth]{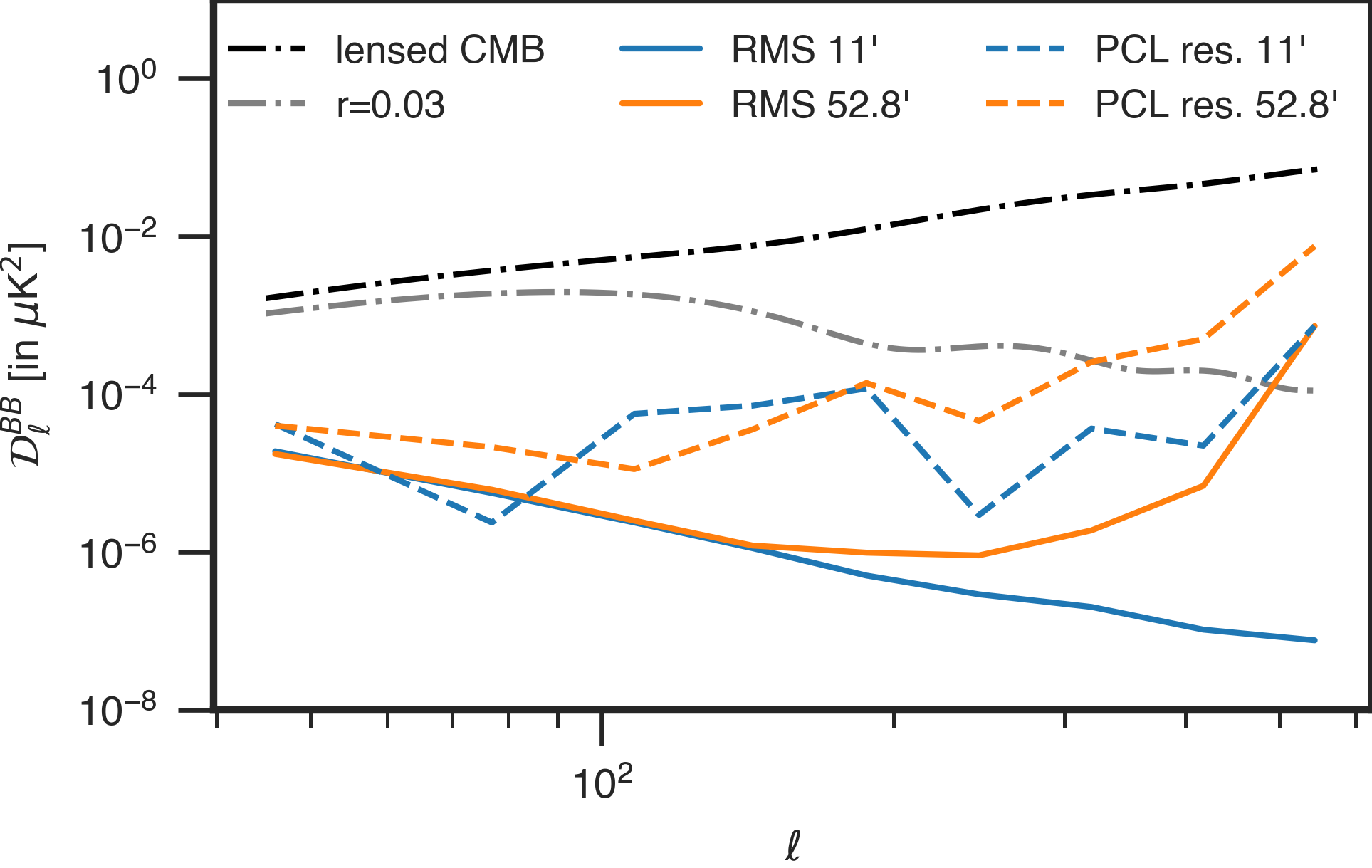}
    \end{subfigure}
    \begin{subfigure}[b]{0.49\textwidth}
        \centering
        \includegraphics[width=\textwidth]{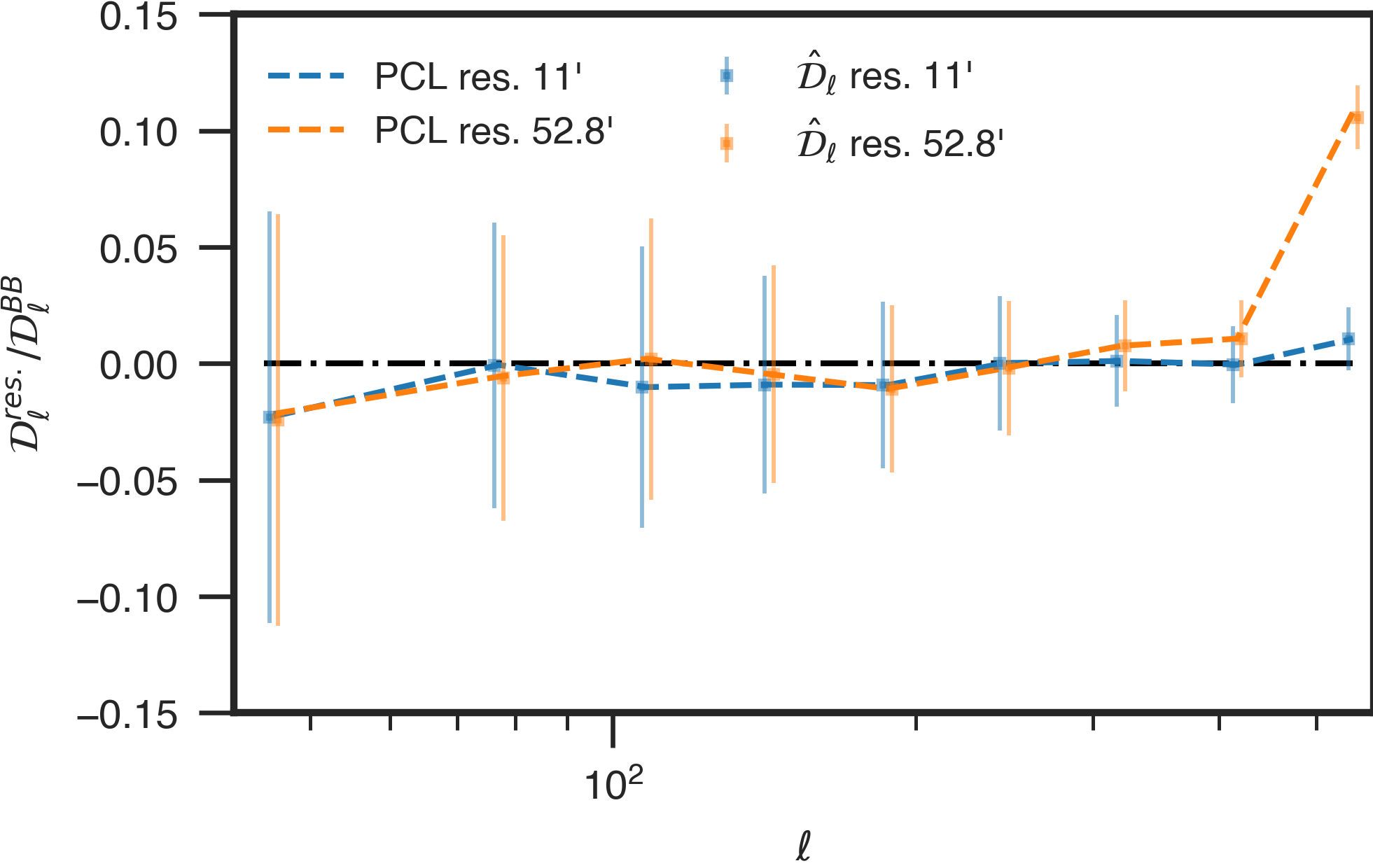}
    \end{subfigure}
    \caption{Left: the RMS defined as Eq.\eqref{eq:rms} and the PCL residual averaged over 100 CMB-only simulations with the beam size of either 11 arcmin or 52.8 arcmin. Right: the ratios between the residual of the estimated power spectrum and the input BB power spectrum. The relative PCL residuals are also shown by dashed curves.}
    \label{fig:pcl-tc}
\end{figure}

To evaluate the residual leakage, we compute the RMS error which depends only on the error of template cleaning as:
\begin{equation}\label{eq:rms}
    \Delta = \sqrt{\frac{1}{N_{\rm sim}}\sum_{i=1}^{N_{\rm sim}}(\hat{\mathcal{D}_\ell}-\mathcal{D}_\ell^{\rm real})^2}\,,
\end{equation}
where $\hat{\mathcal{D}_\ell}$ is the estimated power spectrum for each simulation and $\mathcal{D}_\ell^{\rm real}$ is obtained by running \texttt{NaMASTER} on the real full-sky B maps with the 6$^\circ$ C2 apodized $20\mu$K mask. Besides, the difference between $\mathcal{D}_\ell^{\rm real}$ and the input CMB spectrum $\mathcal{D}_\ell^{\rm BB}$(which are bandpowers in practice) reflects the residual power spectrum only from the PCL estimation. The RMS error (i.e. the error from the residual EB leakage) as well as the mean PCL residual $\langle \mathcal{D}_\ell^{\rm real} - \mathcal{D}_\ell^{\rm BB}\rangle$ are shown in the left panel of Figure~\ref{fig:pcl-tc}. The right panel shows the ratio between the residual of the estimated power spectrum and the input BB power spectrum, $\langle \hat{\mathcal{D}_\ell} - \mathcal{D}_\ell^{\rm BB}\rangle/\mathcal{D}_\ell^{\rm BB}$, averaged over 100 CMB-only simulations. It can be seen that the RMS or the residual leakage is negligible. Thus nearly all the residual comes from the PCL estimator (as shown in the right panel), while the PCL residual surpasses the power spectrum for $r=0.03$ over $\ell\sim300$ in the case of FWHM=52.8 arcmin. The PCL residual might come from the large computational uncertainty of the deconvolved power spectrum after dividing by the beam function squared (see Figure~\ref{fig:need-bands} for the beam functions).
Since the bias of the power spectrum estimation at the beam size of 52.8 arcmin exceeds the 1$\sigma$ threshold over $\ell\sim400$, the multipole range is restricted from 40 to 400 when adopting the common beam of 52.8 arcmin. Hereafter, unless specially mentioned, we adopt FWHM=11 arcmin as the common beam while discarding the K-band data at scales of $\ell>300$ to avoid the extreme noise amplification of this specific frequency channel. 

\subsection{ILC implemention in harmonic domain}
\label{sec:chILC-imp}

Here, we clarify the detailed procedure of foreground cleaning using cILC in harmonic space. After template cleaning, the multi-band B-mode maps masked by the UNP-inv mask are transformed to harmonic space and reconvolved to a common beam. The $\ell$-dependent empirical covariance is averaged by the $a_{\ell m}$s over the bin range $[0.6\ell,1.4\ell]$ with weights of $(2\ell+1)$. We then multiply the cILC weights computed from the covariance matrix by the channel-wise B-mode $a_{\ell m}$s, resulting in the cleaned B-mode $a_{\ell m}$. The final full-sky power spectrum is obtained by PCL after debiasing the noise power spectrum.

\subsection{ILC implemention in needlet (pixel) domain}
\label{sec:cNILC-imp}
\begin{figure}[tbp]
    \centering
    \begin{subfigure}[b]{0.7\textwidth}
        \centering
        \includegraphics[width=\textwidth]{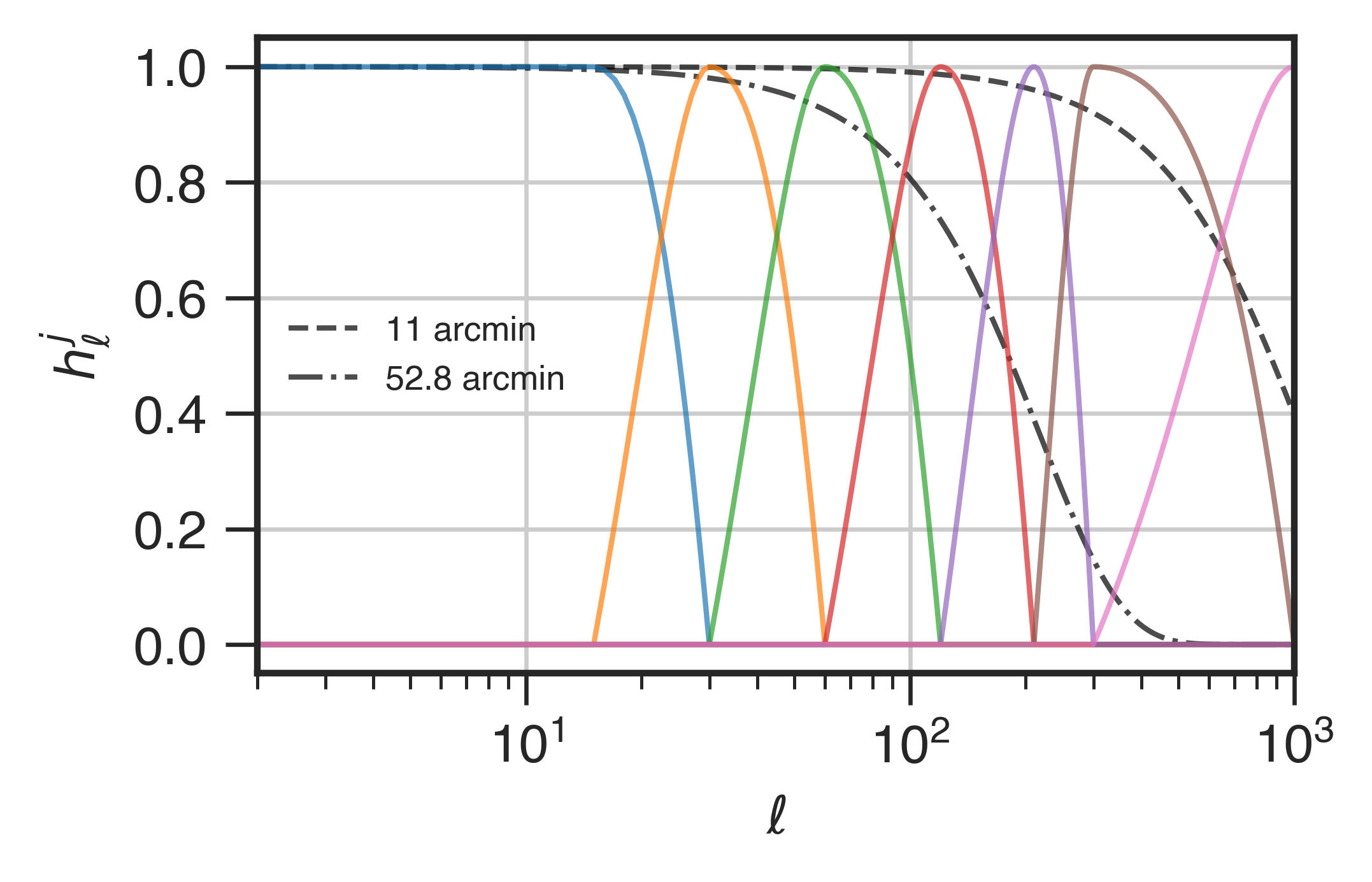}
    \end{subfigure}
    \caption{The window fuctions of seven needlet bands. The beam functions of 11 arcmin and 52.8 arcmin are added for ancillary view.}
    \label{fig:need-bands}
\end{figure}

The ILC implemention in needlet space is somewhat more complicated than in harmonic space due to the discriminatory handling of multi-scale needlet bands. The maps after template cleaning are reconvolved to a common beam and decomposed into needlet coefficients, where the cosine bands are adopted as the needlet bands in this work, as shown in Figure~\ref{fig:need-bands}. The cosine window functions can be written as:
\begin{equation}
    h_\ell^j=\left\{
        \begin{aligned}
           &\cos(\frac{\pi}{2}\frac{\ell_{mid}^j-\ell}{\ell_{mid}^j-\ell_{mid}^{j-1}}) \,, \ell_{mid}^{j-1}\leq\ell<\ell_{mid}^j \,,\\
           &\cos(\frac{\pi}{2}\frac{\ell-\ell_{mid}^j}{\ell_{mid}^{j+1}-\ell_{mid}^j}) \,, \ell_{mid}^j\leq\ell<\ell_{mid}^{j+1} \,,\\
           & 0 \,, \text{otherwise}\,,
        \end{aligned}
    \right.
\end{equation}
where $j\in\{1,2,\dots,n_j\}$, $h_\ell^1=1$ when $0\leq\ell<\ell_{mid}^1$, and $h_\ell^{n_j}=1$ when $\ell\geq\ell_{mid}^{n_j}$. The seven needlet bands used in our tests are listed in Table~\ref{tab:need}. Note that the needlet maps of larger scales containing less modes are therefore downgraded to lower $N_{side}$s. Just like for chILC, we discard K-band data for those needlet bands covering the range of $\ell>350$, i.e. the 6th and 7th bands. The needlet coefficients are also reconvolved with a common beam of 11 arcmin FWHM.

The covariance matrices $\boldsymbol{\hat C}_{jk}$ for the $k$-th pixel of the $j$-th needlet scale are estimated as follows: first we downgrade the $j$-th needlet maps of every channel to a super $N_{side}$ of $N_{side}(j)/4$. Then, the product of downgraded needlets of two channels $\nu_1$ and $\nu_2$, \\ $b^{\nu_1}_j(n_{jk}) b^{\nu_2}_j(n_{jk})$, is smoothed with a Gaussian beam whose FWHM varies with the scale $j$. Each channel pair of needlets returns a covariance map, of which the $k$-th pixel is taken as the $(\nu_1,\nu_2)$ element of the covariance matrix $\boldsymbol{\hat C}_{jk}$. The estimated covariance contains the information of pixels in a disk centered at the $k$-th pixel, averaged by a Gaussian beam. The smoothing FWHMs of seven needlet bands are listed in Table~\ref{tab:need}. The FWHM values are chosen such that the ILC bias \cite{delabrouilleFullSkyLow2009} is under 3\% of the CMB. 

\begin{table}[tbp]
    \centering
    \caption{The needlet bands used for needlet decomposition and the corresponding smoothing FWHMs when computing the covaraince matrix.}
    \begin{tabular}{lccccccc}
        \bottomrule
        Band($j$)&1& 2& 3& 4& 5& 6& 7 \\
        \hline

        $\ell_{mid}^j$&15& 30& 60& 120& 210& 300& 1000 \\
        \hline

        $N_{side}(j)$&32& 64& 128& 256& 512& 1024& 1024 \\
        \hline
        FWHM($^\circ$)&124& 72& 36& 20& 15& 4.2& 3.5 \\
        \toprule
    \end{tabular}
    \label{tab:need}
\end{table}

We compute the weights for each $n_{jk}$ from the covariance matrix independently and project them on the channel-wise needlet maps to obtain the foreground-cleaned needlet maps. Finally we synthesize them to the final cNILC cleaned map, as the inversion of needlet decomposition.

The cILC in pixel space (cpILC) is basically a simple version of cNILC without needlet filtering. Note that in this case $\ell$ is limited up to 300 since the debeamed K band data over the limit would be too large to recover any useful information. As mentioned in Section~\ref{sec:cMILC}, the cMILC is implemented in needlet space adding one more constraint. Thereby, the only processing difference between cMILC and cNILC is that the mixing matrix of cMILC contains an additional column vector describing the SED of the derivative of the thermal dust.

\subsection{Noise debiasing}
It is essential to debias the noise power spectrum from the foreground-cleaned power spectrum where the residual noise power is the predominant component for B-modes. The noise bias can be properly corrected either by the cross spectrum of two data splits between which the noises are uncorrelated, or by estimating the residual noise power spectrum from noise simulations. In this work, we estimate the noise bias from 100 noise simulations projected by the cILC(or NILC) weights obtained by the mock data.

\subsection{Fitting the tensor-to-scalar ratio}

We fit the tensor-to-scalar ratio $r$ with the estimated power spectra by the Markov Chain Monte Carlo (MCMC) simulations. Ignoring the slight non-Gaussianity from the foregrounds given the large number of multipoles in a bin, the Gaussian likelihood of $r$ given the binned cILC cleaned power spectrum $\hat C_{\ell_b}$ is:
\begin{equation}
    - 2 \ln \mathcal{L}(r) = \sum_{\ell_b \ell_{b'}} \left[\hat C_{\ell_b} - rC_{\ell_b}^{r=1} - C_{\ell_b}^{\rm lens} \right]\left[M^{-1}_{\rm fid}\right]_{ \ell_b \ell_{b'}} \left[\hat C_{\ell_{b'}} - rC_{\ell_{b'}}^{r=1} - C_{\ell_b}^{\rm lens} \right]\,,
    \label{eq:gauss_like}
\end{equation}
where $\ell_b, \ell_{b'}$ are indices for the multipole bins, $rC_{\ell_b}^{r=1} + C_{\ell_b}^{\rm lens}$ is the binned theoretical B-mode power spectrum with multipole range $[40, 200]$ including 5 $\ell_b$s, and the fiducial covariance matrix, $[\boldsymbol M_{\rm fid}]_{\ell_b \ell_{b'}}=\langle(\hat C_{{\rm fid,}\ell_b} - \langle \hat C_{{\rm fid,}\ell_b}\rangle) (\hat C_{{\rm fid,}\ell_{b'}} - \langle \hat C_{{\rm fid,}\ell_{b'}}\rangle)\rangle$ is computed from 300 cILC-cleaned simulations with $r=0$ and the foregrounds same with the fitted data. The theoretical spectrum is computed from \texttt{CAMB} with the best-fit parameters of Planck 2018 results \cite{aghanimPlanck2018Results2020a}. We do no delensing to the B-mode power spectrum. The above likelihood is multiplied by a uniform prior of $r\in[0,1]$ to obtain the unnormalized posterior $P(r|\boldsymbol d)$. We set a MCMC chain including 10000 samples satisfying the posterior distribution using the \texttt{emcee} \cite{foreman-mackeyEmceeMCMCHammer2013} python package. Therefore, the mean r-value extracted from the distribution of 10000 samples can be taken as the best-fit $r$ for the estimated power spectrum, with the uncertainty $\sigma(r)$ (also extracted from the distribution) due to the cosmic variance of noise and residual foregrounds embedded in the fiducial covariance matrix.

\section{Results}
\label{sec:results}

In this section, we present the results of the foreground-cleaned maps and the estimated B-mode power spectra using different ILC methods. The input realizations are the sum of CMB, noise and AliCPT foregrounds unless specified. We analyze the possible sources of bias and quantitatively analyze the bias in our results in Section~\ref{sec:bias} and \ref{sec:nbe}. We present the results considering various foregrounds while using the assumed mixing matrix in Section~\ref{sec:res-fg}. We compute the best-fit $r$ and its uncertainty for different cases in Section~\ref{sec:rfit}.
\subsection{Maps}
\begin{figure}[tbp]
    \centering
    \begin{subfigure}[b]{0.32\textwidth}
        \centering
        \includegraphics[width=\textwidth]{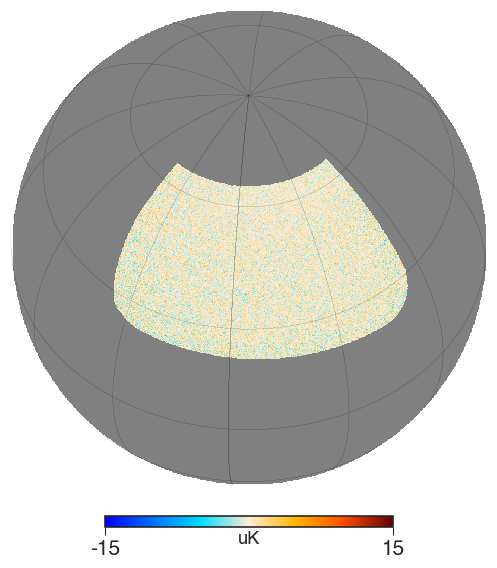}
        \caption{NILC cleaned map}
    \end{subfigure}
    \begin{subfigure}[b]{0.32\textwidth}
        \centering
        \includegraphics[width=\textwidth]{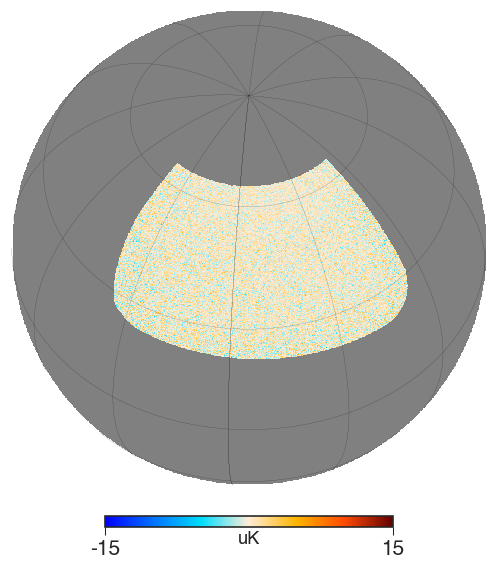}
        \caption{NILC residual noise}
    \end{subfigure}
    \begin{subfigure}[b]{0.32\textwidth}
        \centering
        \includegraphics[width=\textwidth]{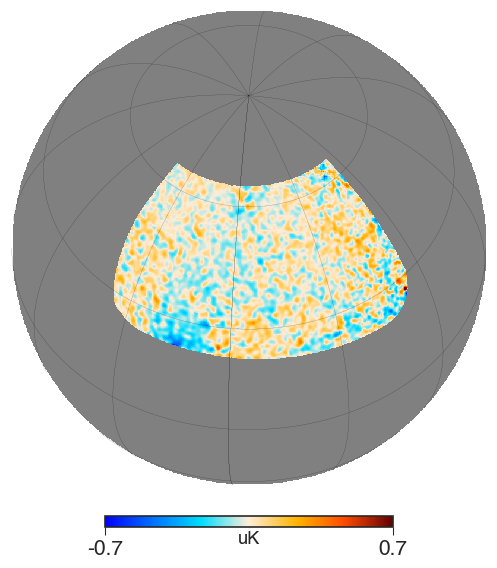}
        \caption{NILC residual foregrounds}
    \end{subfigure}
    \\
    \begin{subfigure}[b]{0.32\textwidth}
        \centering
        \includegraphics[width=\textwidth]{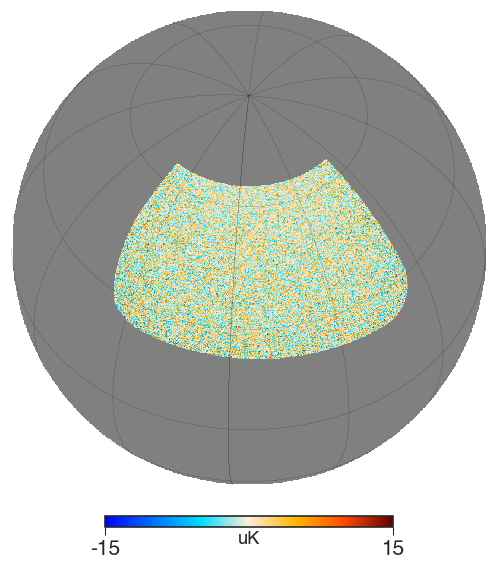}
        \caption{cNILC cleaned map}
    \end{subfigure}
    \begin{subfigure}[b]{0.32\textwidth}
        \centering
        \includegraphics[width=\textwidth]{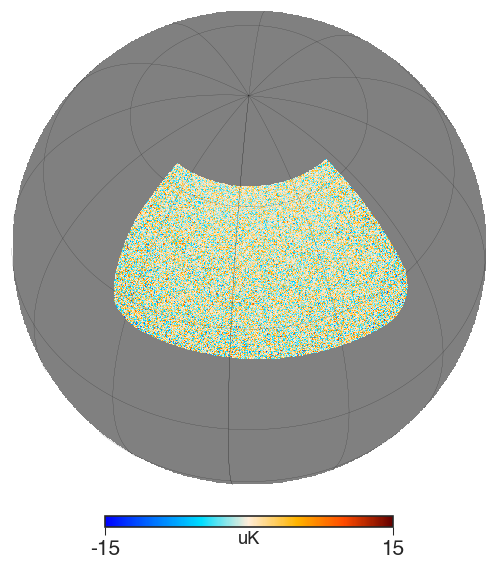}
        \caption{cNILC residual noise}
    \end{subfigure}
    \begin{subfigure}[b]{0.32\textwidth}
        \centering
        \includegraphics[width=\textwidth]{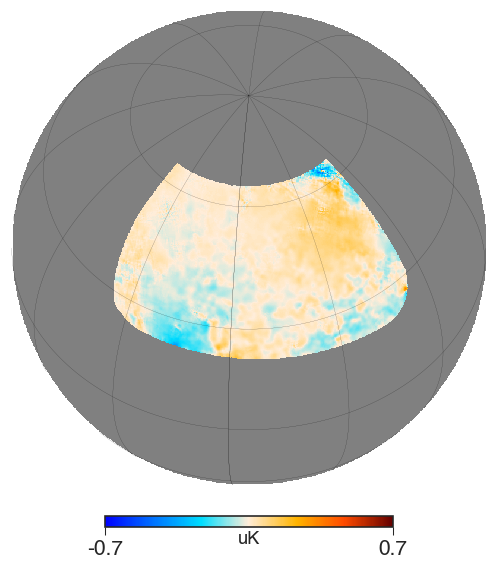}
        \caption{cNILC residual foregrounds}
    \end{subfigure}
    \\
    \begin{subfigure}[b]{0.32\textwidth}
        \centering
        \includegraphics[width=\textwidth]{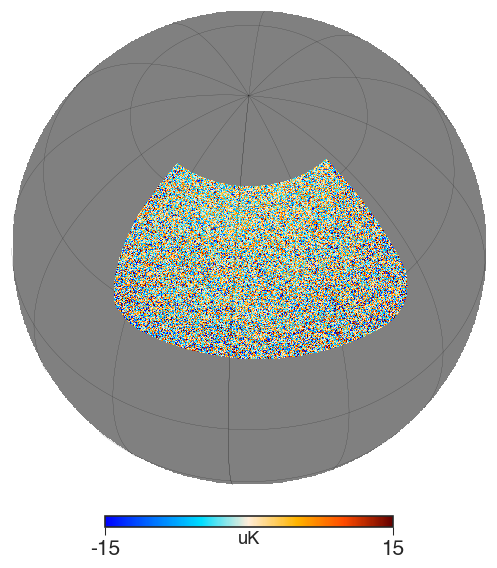}
        \caption{cMILC cleaned map}
    \end{subfigure}
    \begin{subfigure}[b]{0.32\textwidth}
        \centering
        \includegraphics[width=\textwidth]{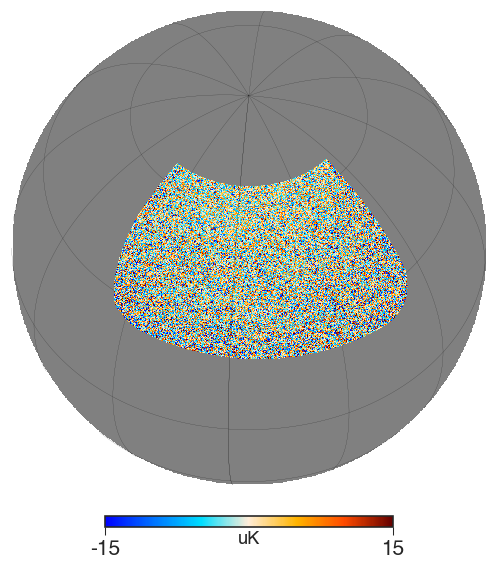}
        \caption{cMILC residual noise}
    \end{subfigure}
    \begin{subfigure}[b]{0.32\textwidth}
        \centering
        \includegraphics[width=\textwidth]{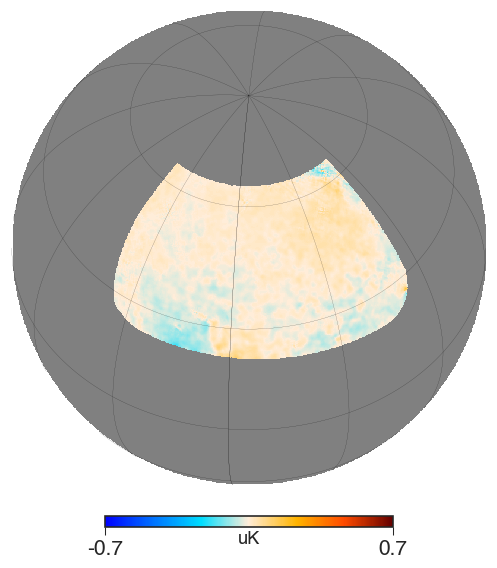}
        \caption{cMILC residual foregrounds}
    \end{subfigure}
    \caption{The cleaned CMB maps, the residual foreground maps and the residual noise maps. From top to bottom are results from NILC, cNILC and cMILC.}
    \label{fig:maps}
\end{figure}

First we illustrate the effectiveness of ILC methods on recovering the CMB map. The cleaned CMB maps and the residual foreground and noise contaminations of NILC, cNILC and cMILC, three ILC methods with increasing number of constraints, are shown in Figure~\ref{fig:maps}. The residual contaminations are produced by applying the same ILC weights on the input foreground and noise maps with the input simulations. It can be seen that the additional constraints on foregrounds improve the foreground removal at the expense of increasing the noise residual in the recovered maps. Which method is the best depends on the object to study. On the map level, adding the constraints introduces much more noise contaminations than the foregrounds further removed. On the power spectrum level, as the noise residual can be properly estimated by noise realizations, it becomes significant to reduce the foreground bias. 

\subsection{Power spectra}
\begin{figure}[tbp]
    \centering
    \begin{subfigure}[b]{0.7\textwidth}
        \centering
        \includegraphics[width=\textwidth]{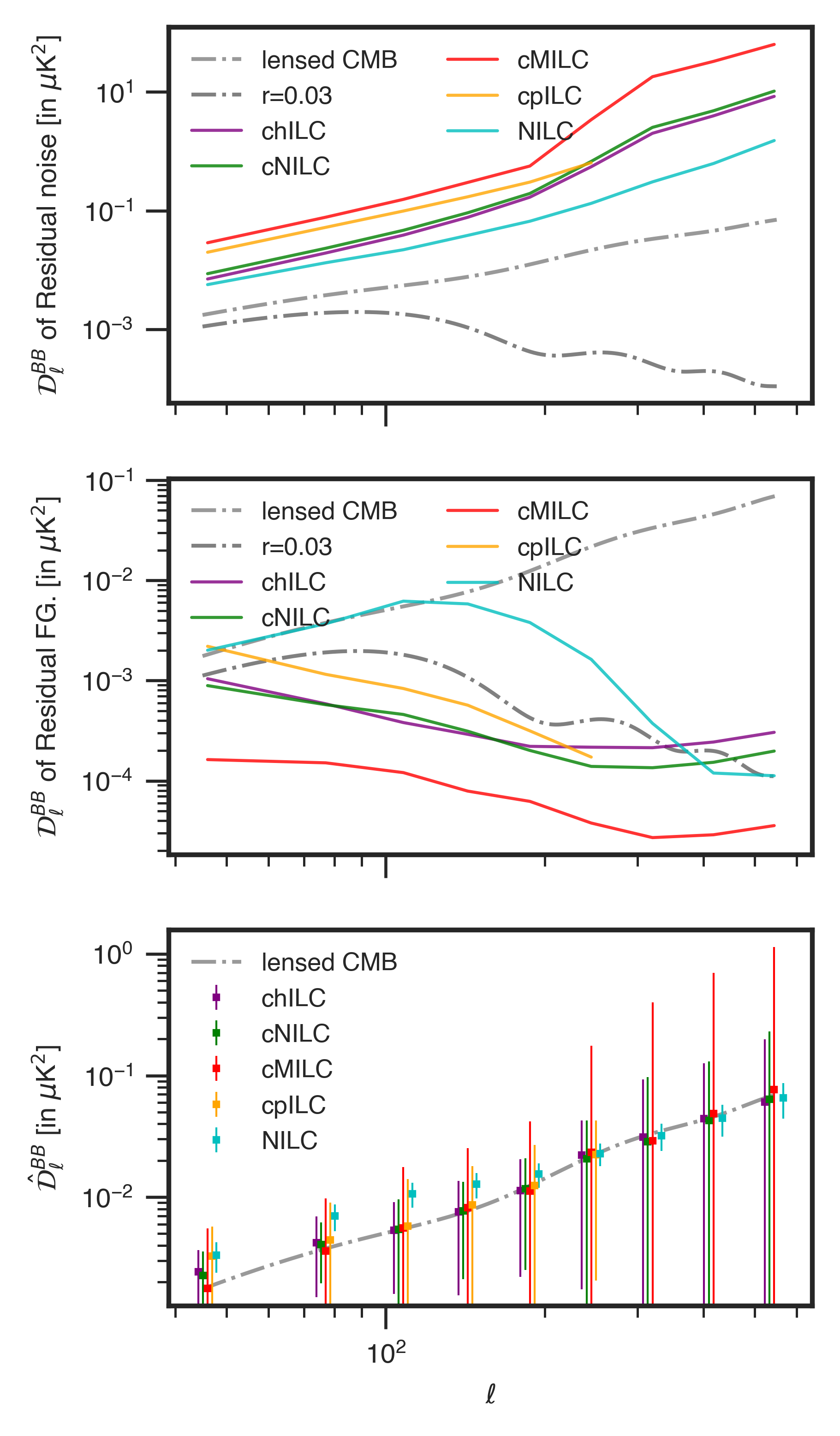}
    \end{subfigure}
    \caption{The comparison between the BB residual noise power spectra (upper), the BB residual foreground power spectra (middle) and the estimated CMB BB power spectra $\hat{\mathcal{D}}_{\ell}^{BB}$ (lower) with different methods. Different colors represent different foreground cleaning methods. The power spectrum of cpILC has a limit of $\ell_{\max}=300$ as mentioned in Section~\ref{sec:cNILC-imp}.}
    \label{fig:Dl}
\end{figure}

The comparison of the estimated power spectra using NILC, chILC, cpILC, cNILC and cMILC and their respective uncertainties are shown in Figure~\ref{fig:Dl}. The effectiveness and the accuracy of foreground cleaning can be quantified in a straightforward way by analyzing both noise and the foreground residuals. They are obtained by linearly combining the input noise  and foregrounds simulations with the ILC weights obtained from the input realization.

It can be seen that compared to NILC, adding constraints reduces the foreground residual significantly while increasing the noise residual since the variance $\boldsymbol{w}^T\boldsymbol{\hat C}\boldsymbol{w}$ with constrained weights cannot remain as small as the unconstrained variance (see Eq.(39) of \cite{remazeillesPeelingForegroundsConstrained2021}). The cMILC adds more constraints than cILC, resulting in the lowest residual foreground at the expense of the largest noise level. Moreover, the performance of the cpILC is poor, meaning that its foreground residual and noise residual are both higher than chILC and cNILC. The chILC result is fairly similar to the cNILC result, except its overall lower noise level and slightly lower foreground residual at scales of $\ell\sim100$.

The ILC method makes a tradeoff between reducing the residual foregrounds and reducing the residual noise. The ILC in harmonic space does not take into account the spatial variation of the noise and the foregrounds, where the noise dominates at high latitude and the foregrounds dominate near the Galactic plane, thus being unfit for use over large sky patches. On the other hand, the ILC in pixel space does not take into account the harmonic variation where the noise dominates at small scales and the foregrounds dominate at large scales. Typically, the ILC in needlet space computes the optimal weights both across the sky and the scales given the localization both in pixel space and in harmonic space. 

In our sky patch, the foreground maps are nearly uniform, which narrows the advantage of localization in both pixel space and favours a fine division in harmonic space. Therefore, the chILC, which assigns weights at each multipole, performs nearly equally well with the cNILC. The cNILC would be expected more advantaged than chILC in cleaning a patch of spatially-varying foregrounds. As for different foreground models in Figure~\ref{fig:4models}, we find that the cNILC does render lower residual foregrounds but a slightly higher noise level in exchange.

Both chILC and cNILC reduces the foregrounds to a low level, which is already low enough. Hence adding yet more constraints with the disadvantage of increasing the noise level might not be a wise choice. For this reason, we will not adopt cMILC in the following tests in view of its large uncertainty of power spectrum estimation. Furthermore, neither NILC nor cpILC will be discussed in the following analysis since they are not able to clean the foregrounds substantially.

\subsection{Bias analysis}
\label{sec:bias}
In this section, we compute analytical estimates of the potential biases of cILC other than the foreground and noise residuals and compute its level in practice. The cILC cleaned map can be decomposed into three components, the CMB signal, the residual foreground and the residual noise:

\begin{equation}\label{eq:slm}
    \hat s_{\ell m}^{\rm cILC}=s_{\ell m}+\boldsymbol{w}^T_{\ell}\boldsymbol{a}^{\rm fg}_{\ell m}+\boldsymbol{w}^T_{\ell}\boldsymbol{n}_{\ell m}\,.
\end{equation}
where the residual noise is the dominant component, while the residual foreground level is of the order $r\sim0.01$ for the tensor spectrum at the large scales. Therefore, it is safe to omit the correlations between the residual foreground and other emission components. Then the power spectrum of the cILC cleaned map is given by:

\begin{equation}
    \hat C_\ell^{\rm cILC}=C^{\rm lens}_\ell+rC^{ r=1}_\ell+C^{\rm res.fg}_\ell+{N}^{\rm res.}_\ell+2C^{\rm s\times res.n}_\ell\,,
\end{equation}
where the chance correlation between CMB and the residual noise $C^{\rm s\times res.n}_\ell$, also denoted as the `ILC bias' \cite{delabrouilleFullSkyLow2009}, can be restricted to a negligible level given enough modes to sample the covariance. The residual noise ${N}^{\rm res.}_\ell$ is debiased by the average of 300 noise realizations weighted by the same cILC weights, denoted by $\bar{N}^{\rm res.}_\ell$. The lensing term $C^{\rm lens}_\ell$ is assumed to be known in this work, with the lensing noise to be studied in the future.

\begin{figure}[tbp]
    \centering
    \begin{subfigure}[b]{0.99\textwidth}
        \centering
        \includegraphics[width=\textwidth]{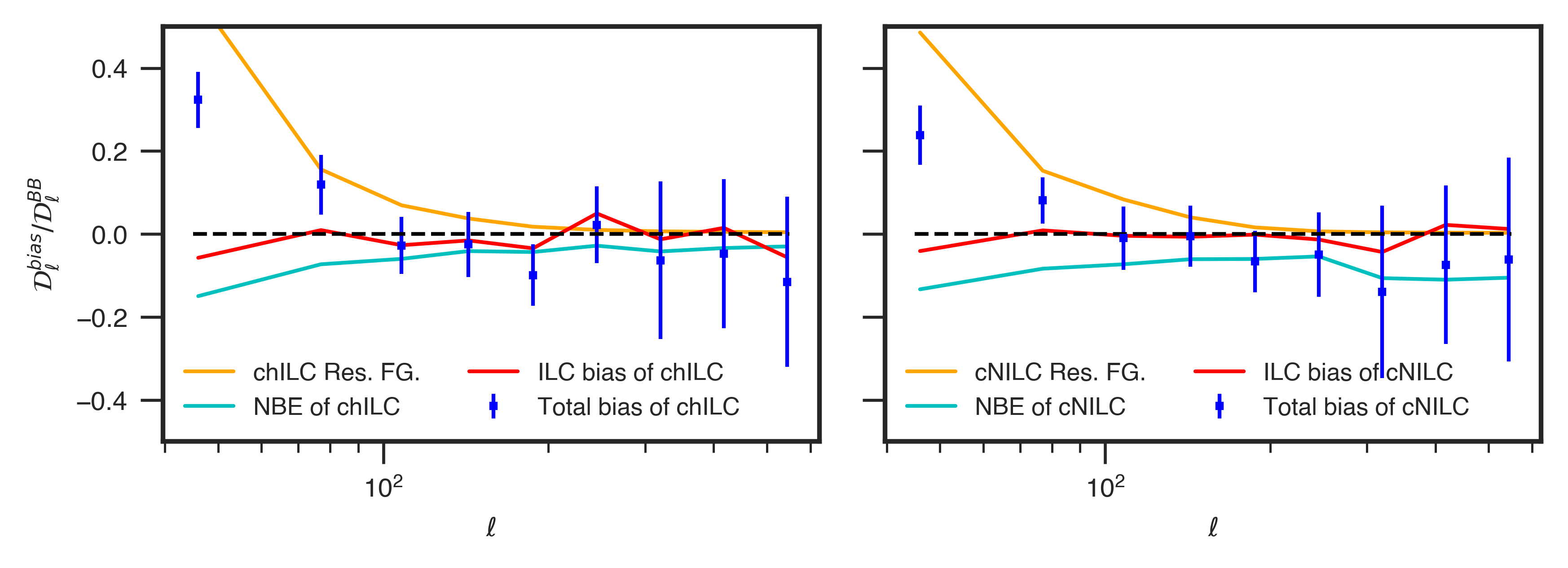}
    \end{subfigure}
    \caption{Ratios of biases from different sources, with the left panel for chILC and the right panel for cNILC. Each panel shows the total bias (blue) with the error bar of $\sigma(\hat{C_\ell})/\sqrt{N_{\rm sim}}$ and its components: the residual foregrounds (orange), the ILC bias (red) and the noise bias error (cyan).}
    \label{fig:all-biases}
\end{figure}

Averaged over realizations, the residual bias of the noise-debiased cILC power spectrum is given by:
\begin{equation}
    R_\ell\equiv\langle\hat C_\ell^{\rm cILC}-\bar{N}^{\rm res.}_\ell-C_\ell^{\rm s}\rangle=\langle C^{\rm res.fg}_\ell\rangle+\langle {N}^{\rm res.}_\ell - \bar{N}^{\rm res.}_\ell \rangle+2\langle C^{\rm s\times res.n}_\ell\rangle\,,
\end{equation}
where $C_\ell^{\rm s}=C^{\rm lens}_\ell+rC^{ r=1}_\ell$ is the input CMB power spectrum and the brakets denote the ensemble averaging. More specifically,
\begin{equation}
    R_\ell=\langle C^{\rm res.fg}_\ell\rangle+\langle \boldsymbol{w}^T_{\ell}(\boldsymbol{N}_\ell - \bar{\boldsymbol{N}}_\ell)\boldsymbol{w}_{\ell}\rangle+\frac{1}{2\ell+1}\sum_{m=-\ell}^{\ell}\langle s^{*}_{\ell m}\boldsymbol{w}^T_{\ell}\boldsymbol{n}_{\ell m}+s_{\ell m}\boldsymbol{w}^T_{\ell}\boldsymbol{n}^{*}_{\ell m}\rangle\,,
\end{equation}
where $\boldsymbol{N}_\ell=\frac{1}{2\ell+1}\sum_{m=-\ell}^{\ell}\boldsymbol{n}_{\ell m}\boldsymbol{n}^{\dagger}_{\ell m}$. The foreground residual can be easily obtained by projecting the input foregrounds on the weights. The second term, named the noise bias error (NBE), is fully discussed in Section~\ref{sec:nbe}. The third term, i.e. the ILC bias, is described theoretically by the following formula:
\begin{equation}
    2\langle C^{\rm s\times res.n}_\ell\rangle=\frac{2(n_c-n_\nu)}{N_k f_{\rm sky}}\langle C^{\rm s}_\ell\rangle\,,
    \label{eq:ilc-bias}
\end{equation}
where $N_k$ is the number of coefficients in the needlet (harmonic) domain as computing the covariance matrix, $f_{\rm sky}\approx5.9\%$\footnote{The effective sky fraction of the inverse noise variance weighted mask is given by $f^{\rm eff}_{\rm sky}=\langle M\rangle^2/\langle M^2\rangle$ where $M$ is the weight of the mask.} is the sky fraction, $n_c=3$ represents the number of constraints, and $n_\nu=7$ is the number of involved channels. In practice, we exclude the $\ell$ ($n_{jk}$) mode from the empirical covariance matrix used to compute the weights of $\ell$ ($n_{jk}$), i.e. $\ell'\neq\ell$ in Eq.\eqref{eq:hilc-wgts}($k'\neq k$ in Eq.\eqref{eq:nilc-cov}), which is commonly used to reduce the ILC bias by avoiding the correlation between the ILC weights and the data sets which the weights are applied on. The actual ratios between the mean ILC bias and the input CMB power spectrum $2\langle C^{\rm s\times res.n}_\ell\rangle/\langle C^{\rm s}_\ell\rangle$ are plotted in Figure~\ref{fig:all-biases}.
As seen in the plot, the actual ILC bias, mostly negative, is of order 1\% of the CMB power spectrum. The deviation of the actual ILC bias from its negative expectation (Eq.\eqref{eq:ilc-bias}) could be attributed to its large uncertainty compared to its small amplitude.

The mean biases of the residual foreground and the noise bias error are also obtained from simulations. We compute their ratios to the input CMB power spectrum, as shown in the right panel of Figure~\ref{fig:all-biases}. We also plot the total biases $\langle\hat C_\ell - C_\ell^{s}\rangle/C_\ell^{s}$, whose error bars are given by the uncertainty of the mean residual bias $\sigma(\hat{C_\ell})/\sqrt{N_{\rm sim}}=\sigma(\hat{C_\ell})/10$.
As seen in the plot, the biases for chILC and cNILC are mainly from NBE at small scales of $\ell$>100, and from residual foregrounds at large scales. Since the NBE and the ILC bias are both within the uncertainty of the estimated noise bias at all scales, we do not consider it necessary to debias them.

\subsubsection{Noise bias error}
\label{sec:nbe}
In our analysis of the bias, we find that the estimated noise bias is systematically lower than the actual noise bias (obtained from the actual noise projected by the cILC weights) due to the fact that the cILC weights are conceptually designed to minimize the power spectrum of the actual data which is dominated by the actual noise. The error, however, is rarely mentioned by the previous works of ILC either because they use the cross spectra of data splits to cancel the noise bias or their simulations are not dominated by the noise. To make it clear, the third term of Eq.\eqref{eq:slm} is the residual noise, whose power spectrum is dubbed the noise bias ${N}^{\rm res.}_\ell=\boldsymbol{w}^T_{\ell}\hat{\boldsymbol{N}}_\ell\boldsymbol{w}_{\ell}$, while the estimated noise bias averaged over noise simulations is $\bar{N}^{\rm res.}_\ell=\boldsymbol{w}^T_{\ell}\bar{\boldsymbol{N}}_\ell\boldsymbol{w}_{\ell}$. Since the noise covariance dominates the total covariance matrix, as the weights (dependent on the realization) varying, the minimum point of $\boldsymbol{w}^T_{\ell}\boldsymbol{\hat C}_\ell\boldsymbol{w}_{\ell}$ is approximately the minimum point of $\boldsymbol{w}^T_{\ell}\hat{\boldsymbol{N}}_\ell\boldsymbol{w}_{\ell}$; thus the cILC weights also nearly minimizes the noise bias of the particular noise realization, i.e. $\boldsymbol{w}^T_{\ell}\hat{\boldsymbol{N}}_\ell\boldsymbol{w}_{\ell}\leq\boldsymbol{w}^T_{\ell}\bar{\boldsymbol{N}}_\ell\boldsymbol{w}_{\ell}$ for almost any noise realization. The difference between the actual and estimated noise bias, named the noise bias error, is approximated (See Appendix \ref{sec:deriv_nbe} for the derivation) as:
\begin{equation}
    \langle{N}^{\rm res.}_\ell-\bar{N}^{\rm res.}_\ell\rangle
    =
    \langle \boldsymbol{w}^T_{\ell}(\hat{\boldsymbol{N}}_\ell - \bar{ \boldsymbol{N}}_\ell)\boldsymbol{w}_{\ell}\rangle
    \approx
    \frac{2(n_c-n_\nu)}{N_kf_{\rm sky}}\bar{N}^{\rm res.}_\ell\,.
    \label{eq:nbr}
\end{equation}
In our application, $n_c=3$, $n_\nu=7$ at $\ell<300$ or $6$ otherwise, the effective sky fraction of the UNP-inv mask $f_{\rm sky}\approx5.9\%$ and $N_k$ is the number of coefficients involved when computing the covariance matrix in an cILC implementation, e.g., $N_k\sim14000$ at $\ell\sim100$ for the chILC. Typically we use larger $N_k$ to estimate the covariance matrix at smaller scales.

\begin{figure}[tbp]
    \centering
    \begin{subfigure}[b]{0.7\textwidth}
        \centering
        \includegraphics[width=\textwidth]{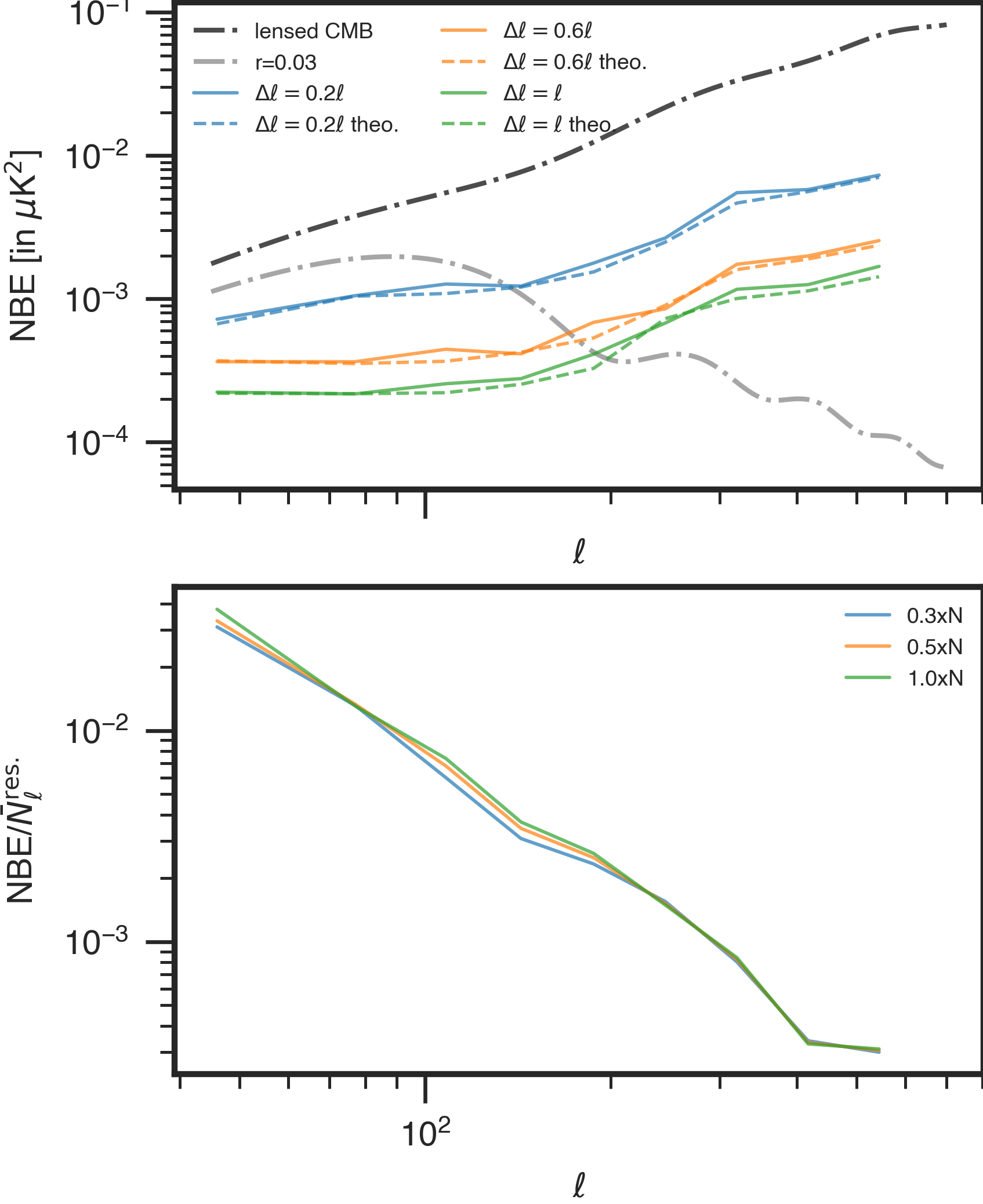}
    \end{subfigure}
    \caption{Upper: The noise bias errors from simulations (solid) compared to their theoretical expectations (dashed) in chILC. Lower: The ratios between the noise bias error and the mean residual noise under different noise levels (0.3, 0.5 and 1.0 times the real noise level) in chILC.}
    \label{fig:nbe}
\end{figure}

We compute the noise bias estimated from 100 noise simulations and the actual noise power spectrum for the chILC using both different binning sizes of the covariance matrix and different noise levels. In chILC, the covariance matrix $\hat{\boldsymbol{C}}_\ell$ is computed by averaging the covariance over harmonic modes in $\ell\pm\Delta\ell/2$. The noise bias error and its theoretical expectations for chILC are plotted in Figure~\ref{fig:nbe}. Being close to the theoretical results, the noise bias error reduces with the binning size (or $N_k$) increasing.

As seen in the lower plot, the ratio between the noise bias error and the mean residual noise does not vary with the noise level, as expected from Equation~\ref{eq:nbr}. The noise bias error is about 0.8\% of the mean residual noise at $\ell\sim100$, which is negligible compared to the uncertainty of the noise bias $\sigma({N}^{\rm res.}_\ell)/\sqrt{N_{\rm sim}}\approx\sqrt{\frac{1}{(2\ell+1)\Delta\ell_b f_{\rm sky}N_{\rm sim}}}\bar{N}^{\rm res.}_\ell$ where $\Delta\ell_b$ is the binning size in the \texttt{NaMASTER} power spectrum estimation and $N_{\rm sim}$ is the number of noise simulations used to debias the noise.

\revise{
We investigate possible ways to mitigate the ILC noise bias error.
Theoretically, either fixing the weights independent of the realizations, or crossing two different data splits to null the average noise bias should eliminate this kind of error. 
Here we use the Jackknife method, where the input maps are two data splits with the same CMB and foregrounds but different Gaussian instrumental noises. We first obtain the cILC weights from the average of two data splits, then apply the weights onto two splits individually and compute their cross power spectrum to get rid of the noise bias. The results shown in Appendix \ref{sec:ver_nbe} have verified that the Jackknife would not produce the noise bias error.
}

\subsection{Other foreground models}
\label{sec:res-fg}

\begin{figure}[tbp]
    \centering
    \begin{subfigure}[b]{0.99\textwidth}
        \centering
        \includegraphics[width=\textwidth]{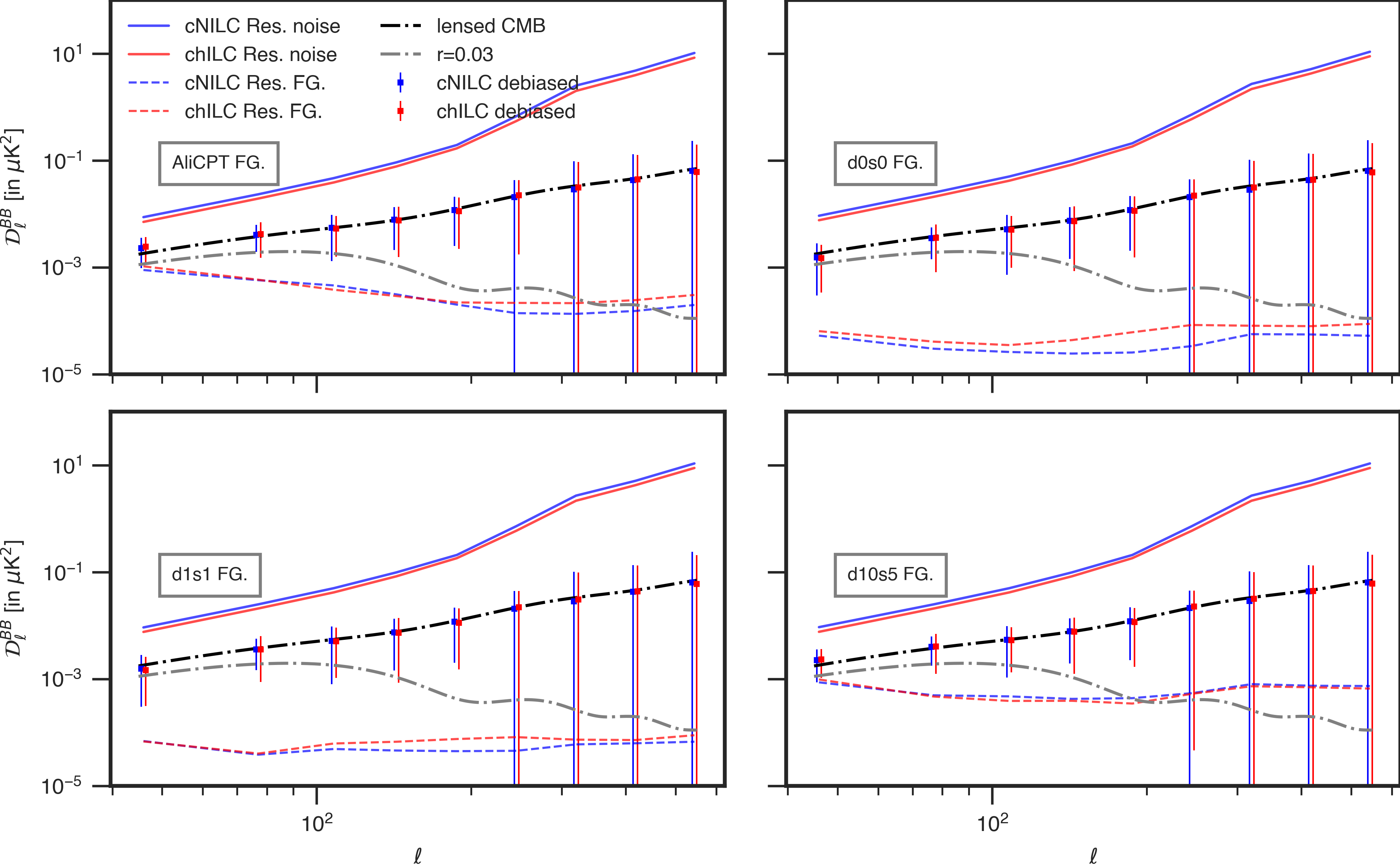}
    \end{subfigure}
    \caption{Results of chILC (red) and cNILC (blue) debiased power spectra and the residual noise and foreground spectra for four different foreground models. The noise residuals are plotted as solid curves and the foreground residuals are plotted as dashed curves.}
    \label{fig:4models}
\end{figure}

We apply chILC and cNILC on four sets of simulations with four types of foregrounds introduced in Section~\ref{sec:fg} to test the robustness of the mixing matrix. As seen in Figure~\ref{fig:4models}, both methods are able to reduce all types of foregrounds effectively. The noise residual hardly varies among models since the data share the same noise realizations, while the foreground residual depends on the particular foreground model. The new PySM synchrotron model \texttt{s5} has more power than the older models \texttt{s0} and \texttt{s1}. The PySM dust model \texttt{d10} has more decorrelation than either \texttt{d0} or \texttt{d1} dust models. The larger residual foregrounds for \texttt{d10s5} compared to \texttt{d0s0} and \texttt{d1s1} likely originates in the complexity of the \texttt{d10s5} model. 

\subsection{Estimation of the tensor-to-scalar ratio}
\label{sec:rfit}
\begin{figure}[tbp]
    \centering
    \begin{subfigure}[b]{0.8\textwidth}
        \centering
        \includegraphics[width=\textwidth]{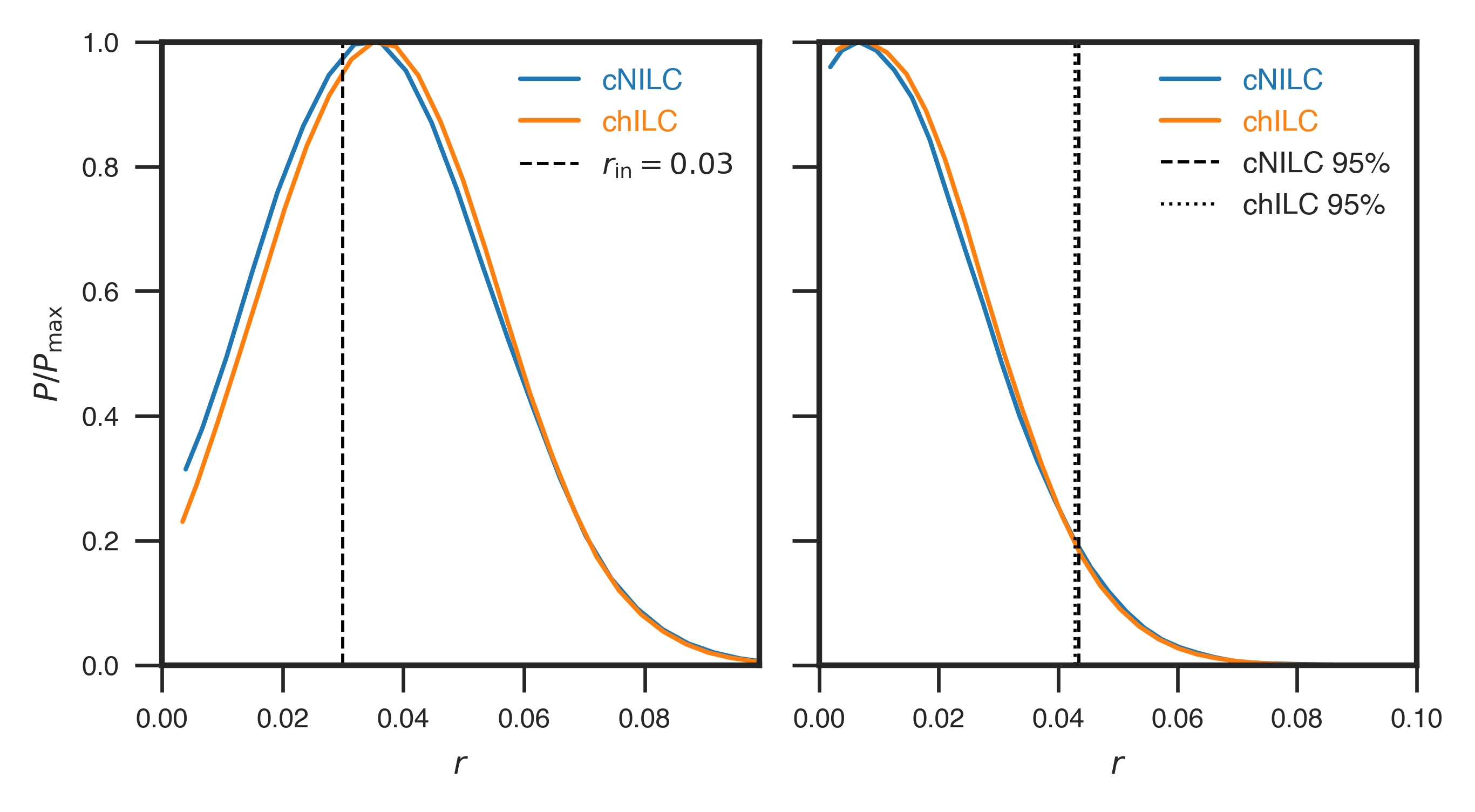}
    \end{subfigure}
    \\
    \begin{subfigure}[b]{0.7\textwidth}
        \centering
        \includegraphics[width=\textwidth]{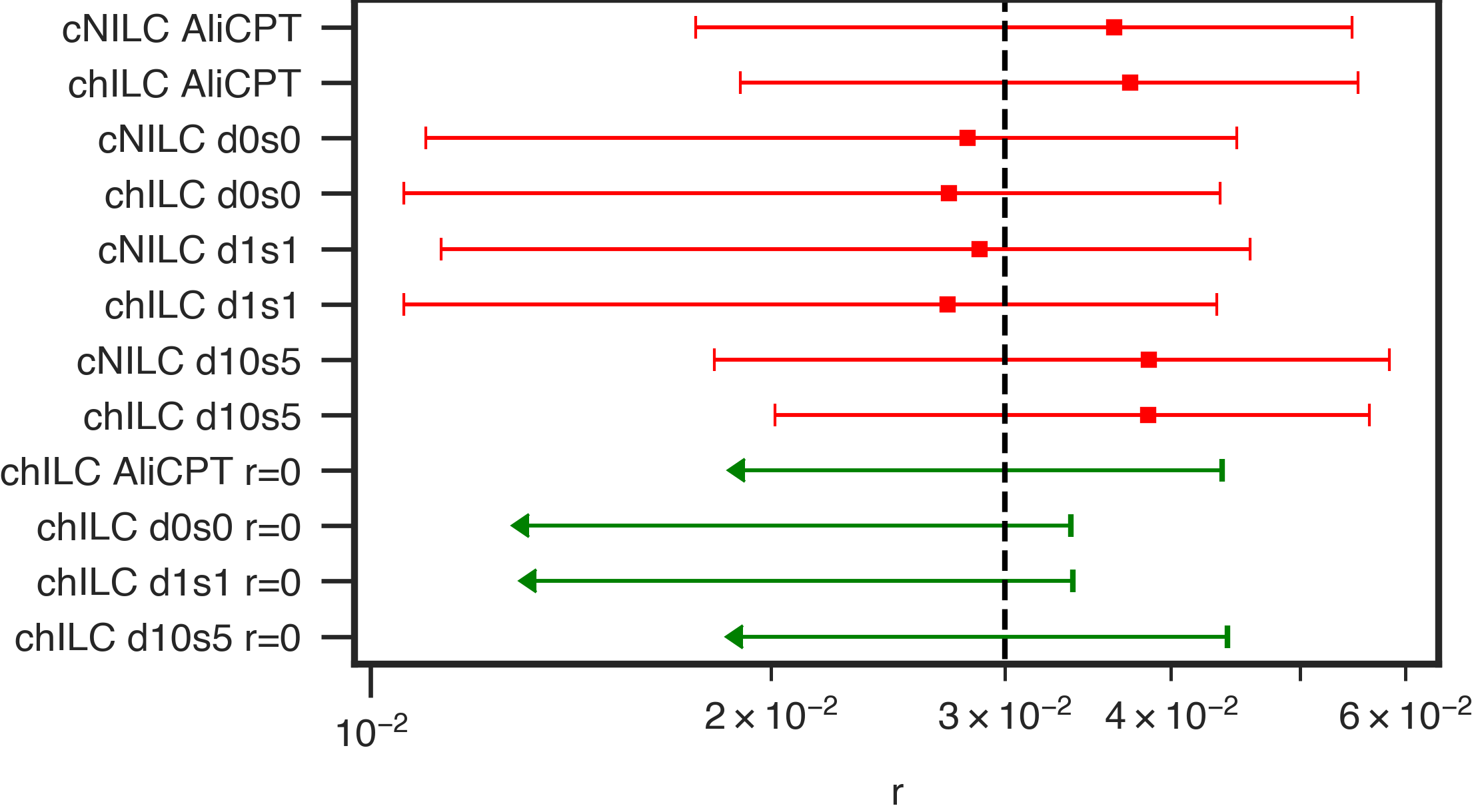}
    \end{subfigure}
    \caption{Left upper: The posterior distribution of $r$ for both chILC (orange) and cNILC (blue) tests with the input value $r=0.03$ and AliCPT foregrounds. Right upper: The posterior distribution of $r$ for both chILC and cNILC null tests with the input $r=0$ and AliCPT foregrounds. Lower: The best-fit tensor-to-scalar ratio $r$ of chILC and cNILC results with the four models considered here. For the input $r=0.03$, the best-fit values for $r$ are shown in red squares with 1$\sigma$ error-bars. \revise{For the null tests with input $r=0$, the best-fit $r$ values are shown in green arrows with 2$\sigma$ upper limits in gree bars, and only the chILC results are shown since they are close to the cNILC results.}}
    \label{fig:likelihood}
\end{figure}

We estimate the tensor-to-scalar ratio $r$ from the mean measured power spectrum after foreground cleaning using chILC and cNILC on our simulated data with four different foreground models. The results are summerized in Figure~\ref{fig:likelihood}. The best-fit $r$ values and the corresponding 1$\sigma$ error bars are shown in red, with the input value $r=0.03$. We can see that the true value $r=0.03$ is within 1$\sigma$ error bars for all cases. For AliCPT foregrounds we obtain $r=0.037\pm0.018$ ($r=0.036\pm0.018$) for chILC (cNILC). For the null tests with the input of $r=0$, the chILC results of best-fit $r$ and 2$\sigma$ upper limits are shown in green arrows and bars respectively, which are similar to the cNILC results. The results are also summarized in Table~\ref{tab:r}. In the null tests of the AliCPT foregrounds, the 95\% CL upper limit of $r$ is 0.043 for both chILC and cNILC. \revise{With the simpler foreground models like \texttt{d0s0} and \texttt{d1s1}, the constraint could reach $r<0.033$ due to the lower foreground bias.}

\begin{table}[tbp]
    \centering
    \caption{Upper: the best-fit $r$ and its uncertainties for the mean cILC cleaned power spectrum with the input $r=0.03$. Lower: the $2\sigma$ constraints (95\%CL) on $r$ for the null test without the tensor B modes.}

    \begin{tabular}{l|c|c}
        \bottomrule
        $r_{\rm in}=0.03$&chILC&cNILC\\
        \hline
        AliCPT
        &$r=0.037\pm0.018$&$r=0.036\pm0.018$\\
        \hline
        d0s0
        &$r=0.027\pm0.016$&$r=0.028\pm0.017$\\
        \hline
        d1s1
        &$r=0.027\pm0.016$&$r=0.029\pm0.017$\\
        \hline
        d10s5
        &$r=0.038\pm0.018$&$r=0.038\pm0.018$ \\
        \toprule
    \end{tabular}
    \begin{tabular}{l|c|c}
        \bottomrule
        Null Test($r_{\rm in}=0$)&chILC&cNILC\\
        \hline
        AliCPT
        &$r<0.043$
        &$r<0.043$  \\
        \hline
        d0s0
        &$r<0.033$
        &$r<0.034$  \\
        \hline
        d1s1
        &$r<0.034$
        &$r<0.034$  \\
        \hline
        d10s5
        &$r<0.044$
        &$r<0.044$  \\
        \toprule
    \end{tabular}
    \label{tab:r}
\end{table}

\section{Conclusions}
\label{sec:conclusions}

The precise measurement of the CMB B-mode power spectrum as well as the tensor-to-scalar ratio demands an elaborate component separation method in order to balance the minimization of the statistical uncertainty and the systematic bias. Without many priors on the Galactic foregrounds, the ILC method and its variants are competent candidates for the future ground-based CMB experiments.

We evaluate the foreground cleaning performance of some variants of the ILC method on partial sky B-modes. We carry out NILC, cMILC, and the cILC methods in different domains (cpILC, chILC, cNILC) on a northern sky patch using the AliCPT, WMAP-K and Planck HFI combined mock data. We mainly focus on the cILC methods in harmonic (chILC) and needlet (cNILC) domain since the NILC results show larger foreground bias. According to our results, there are advantages in different aspects when we use both chILC and cNILC. 

In terms of the power spectrum, the cILC in harmonic space renders slightly less noise residuals and smaller uncertainties on estimation of the power spectrum and the tensor-to-scalar ratio. The cNILC method renders less foreground residuals, thus reducing the systematic errors. The results of both methods are similar due to the relatively low and uniform foreground contamination in our sky patch.  On the other hand, the foreground cleaning performance of the cILC in pixel space is poor due to the properties of the data. Moreover, using cMILC in the AliCPT patch results in high noise uncertainty, despite lowest foreground residuals.

Additionally, doing ILC in harmonic space is about two times computationally faster than that in needlet space, since the size of the covariance matrix for harmonic ILC is smaller on average. Applying the NILC costs plenty of computational hours on needlet transformation. Our results using different foregrounds also demonstrate that our mixing matrix is compatible to various foreground models.

Analysis of main biases on power spectra shows that other than the foreground residual and the ILC bias, the estimated noise bias cannot accurately counteract the actual noise bias given the large noise level and the finite size of the domain where we compute the covariance. Fortunately, this error (named the noise bias error) could be limited to a negligible order compared to the uncertainty of the estimated power spectrum. We firstly derive its relation with respect to the noise residual which fits well with the results from simulations. \revise{We also illustrate that the Jackknife approach crossing two data splits would not produce this error.}

We estimate the tensor-to-scalar ratio $r$ for our estimated power spectra, concluding that the methods bias $r$ in about 0.008 at maximum for all cases with an uncertainty no larger than 0.018. For the null tests without primordial B-modes, we obtain a constraint of $r<0.043\,$(95\%CL) applying cNILC/chILC on AliCPT-1 mock data, and $r<0.033\,$(95\%CL) using the simpler \texttt{d0s0} sky model. Note that since currently we have not yet involved all possible systematics, the constraints might be underestimated and could not be directly compared to the best limit $r<0.034\,$(95\%CL) of pure CMB experiments from combined Planck PR4 + BK18 data, or $r<0.032\,$(95\%CL) with BAO data involved further \cite{tristramImprovedLimitsTensortoscalar2022}. Nevertheless, the future observations with lower noise would be promising to further constrain the results.

\appendix
\section{Effects of point sources}
\label{sec:ps}

\begin{figure}[tbp]
    \centering
    \begin{subfigure}[b]{0.67\textwidth}
        \centering
        \includegraphics[width=\textwidth]{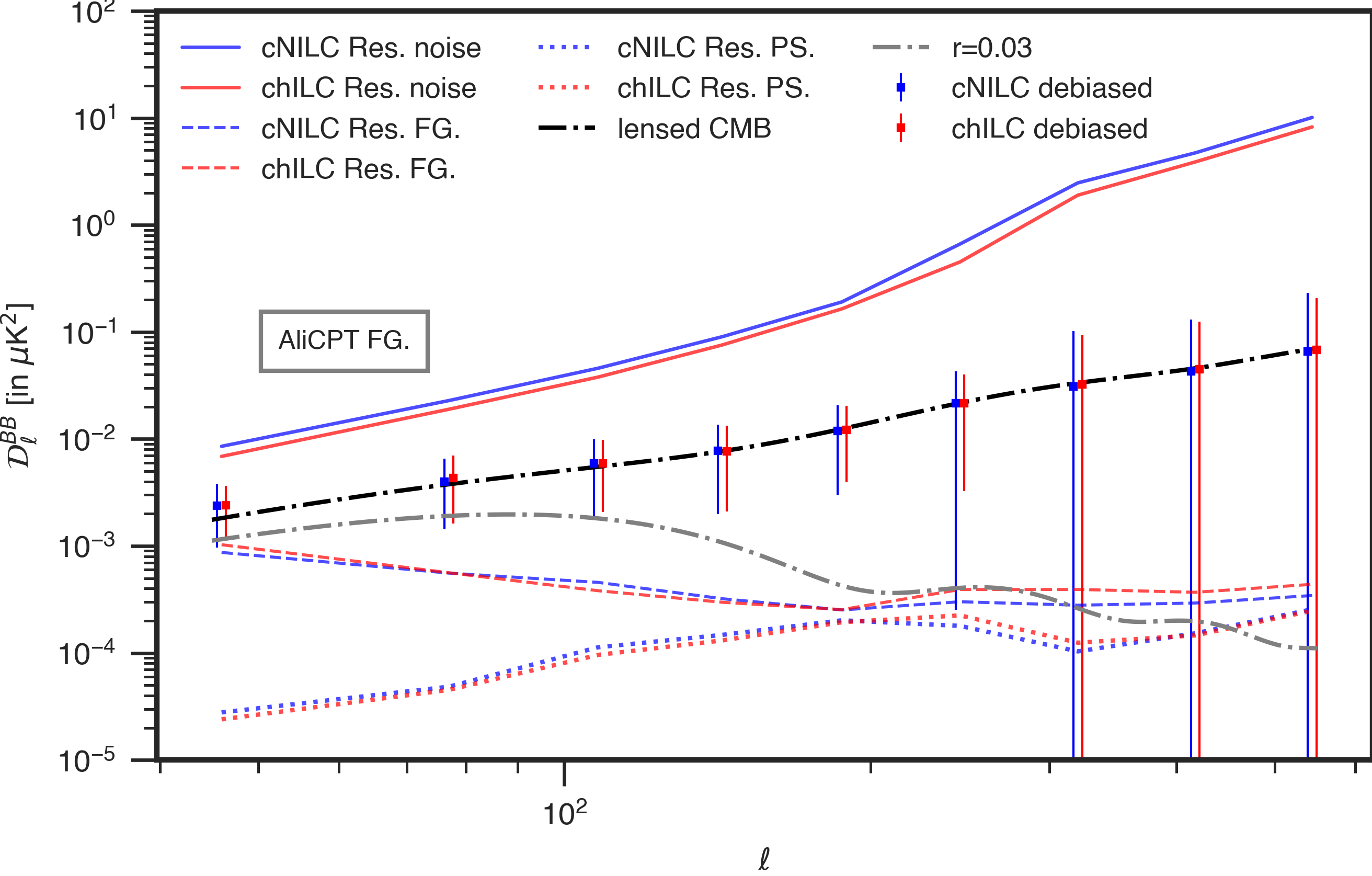}
    \end{subfigure}
    \begin{subfigure}[b]{0.32\textwidth}
        \centering
        \includegraphics[width=\textwidth]{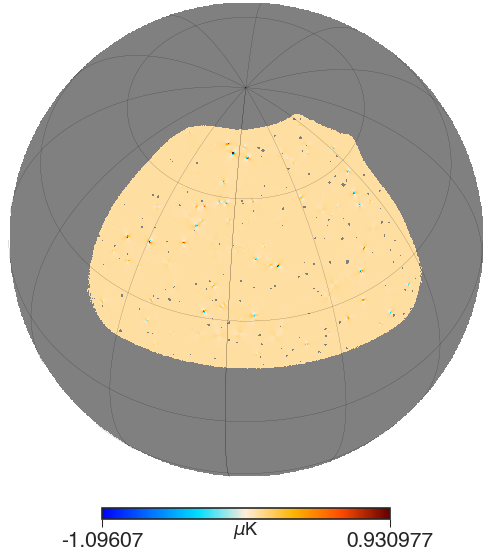}
    \end{subfigure}
    \caption{Testing chILC and cNILC methods considering the point sources. Left: the debiased power spectra and the noise, foreground and point source residuals. Right: the masked point source B-mode map at 150GHz after template cleaning.}
    \label{fig:ps}
\end{figure}

Here, we add a simulated polarized point source map to the data sets. The input point sources are modeled based on the radio source catalogue of the Planck Sky Model \cite{delabrouillePrelaunchPlanckSky2013}. These point sources are masked by a point source mask produced by the matched filter and inpainted before the cILC implementation. Since the point sources are independent of other components, we compute the residual point source power spectra from cNILC and chILC methods by projecting the input point source maps on the cILC weights, as shown in Figure~\ref{fig:ps}. It can be seen that the contribution of residual point sources is negligible compared to the bias of residual foregrounds.

 We preprocess the input maps to diminish the effects of point sources. First, we identify the point sources on the HFI 100GHz temperature map using the outlier method \cite{tegmarkRemovingPointSources1998} and obtain the point source mask using the matched filter with a threshold of 3$\sigma$. Then we mask the QU maps with the point source mask and use the template cleaning method to obtain the B-mode maps with the point sources masked. Finally, we inpaint the B maps using the \texttt{iSAP} \cite{starckISAPInteractiveSparse2013} package. The procedure after point source preprocessing is the same with the basic B-mode reconstruction pipeline.

\section{Derivation of the noise bias error}
\label{sec:deriv_nbe}
In this appendix we compute the analytical form of the noise bias error, which must be carefully accounted for under cases of low signal-to-noise ratio. The derivation is in harmonic space while the result is general in all domains.

\subsection{The noise bias error in standard ILC}
The input data is modelled as $\boldsymbol d_{\ell m}=\boldsymbol a s_{\ell m}+\boldsymbol f_{\ell m}+\boldsymbol n_{\ell m}$ where $s_{\ell m}$ is the CMB signal, $\boldsymbol f_{\ell m}$ the foregrounds and $\boldsymbol n_{\ell m}$ the noise term. Ignoring the binning scheme till the final form of the noise bias error, the estimated covariance matrix (in harmonic ILC) at $\ell$ is given by:
\begin{equation}
    \hat{\boldsymbol C}_{\ell}=\frac{1}{2\ell+1}\sum_{m=-\ell}^{\ell} \boldsymbol d_{\ell m} \boldsymbol d_{\ell m}^\dagger=\hat{C}^s_{\ell}\boldsymbol a\boldsymbol a^T+\hat{\boldsymbol N}_{\ell}+\hat{\boldsymbol F}_\ell+\boldsymbol{\Lambda}_\ell \,,
    \label{eq:c^hat}
\end{equation}
where
\begin{equation}
    \hat{C}^s_{\ell}=\frac{1}{2\ell+1}\sum_{m=-\ell}^{\ell} s_{\ell m} s_{\ell m}^* \,,
\end{equation}
\begin{equation}
    \hat{\boldsymbol N}_{\ell}=\frac{1}{2\ell+1}\sum_{m=-\ell}^{\ell} \boldsymbol n_{\ell m} \boldsymbol n_{\ell m}^\dagger \,,
\end{equation}
\begin{equation}
    \hat{\boldsymbol F}_{\ell}=\frac{1}{2\ell+1}\sum_{m=-\ell}^{\ell} \boldsymbol f_{\ell m} \boldsymbol f_{\ell m}^\dagger \,,
\end{equation}
\begin{equation}
    \boldsymbol{\Lambda}_\ell=\frac{1}{2\ell+1}\sum_{m=-\ell}^{\ell} (s_{\ell m}\boldsymbol a \boldsymbol n_{\ell m}^\dagger+s_{\ell m}^*\boldsymbol n_{\ell m}\boldsymbol a^T) \,.
\end{equation}
We ignore the cross terms between the foregrounds and the CMB or noise. Hereafter, the `hat' over some quantity denotes the empirical estimation from one realization while the `bar' denotes the ensemble average. Then the average terms
\begin{equation}
    \bar{\boldsymbol C}_{\ell}=\frac{1}{2\ell+1}\sum_{m=-\ell}^{\ell} \langle\boldsymbol d_{\ell m} \boldsymbol d_{\ell m}^\dagger\rangle=\bar{C}^s_{\ell}\boldsymbol a\boldsymbol a^T+\bar{\boldsymbol N}_{\ell}+\bar{\boldsymbol F}_{\ell} \,,
    \label{eq:c^bar}
\end{equation}
\begin{equation}
    \bar{C}^s_{\ell}=\frac{1}{2\ell+1}\sum_{m=-\ell}^{\ell} \langle s_{\ell m} s_{\ell m}^*\rangle \,,
\end{equation}
\begin{equation}
    \bar{\boldsymbol N}_\ell=\frac{1}{2\ell+1}\sum_{m=-\ell}^\ell \langle\boldsymbol n_{\ell m} \boldsymbol n_{\ell m}^\dagger\rangle\,,
\end{equation}
\begin{equation}
    \bar{\boldsymbol F}_\ell=\frac{1}{2\ell+1}\sum_{m=-\ell}^\ell \langle\boldsymbol f_{\ell m} \boldsymbol f_{\ell m}^\dagger\rangle\,,
\end{equation}
where $\bar{\boldsymbol{\Lambda}}_\ell=0$ assuming the CMB is not correlated to the noise.

The (harmonic) ILC weights
\begin{equation}
    \hat{\boldsymbol w}^T=\frac{\boldsymbol a^T\hat{\boldsymbol C}_{\ell}^{-1}}{\boldsymbol a^T\hat{\boldsymbol C}_{\ell}^{-1}\boldsymbol a}
\end{equation}
are applied on the $\ell$-mode input data sets.

Given the residual noise power spectrum ${N}^{\rm res.}_\ell=\hat{\boldsymbol w}^T\hat{\boldsymbol N}_\ell\hat{\boldsymbol w}$ and the estimated noise bias $\bar{N}^{\rm res.}_\ell=\hat{\boldsymbol w}^T\bar{\boldsymbol N}_\ell\hat{\boldsymbol w}$, the noise bias error is defined as the average of their difference:
\begin{equation}
    \langle{N}^{\rm res.}_\ell-\bar{N}^{\rm res.}_\ell\rangle=\langle\hat{\boldsymbol w}^T(\hat{\boldsymbol N}_\ell-\bar{\boldsymbol N}_\ell)\hat{\boldsymbol w}\rangle\,.
\end{equation}
Assuming the noise realizations are independent among channels:
\begin{equation}
    \bar{N}_{\ell,ij}=\frac{1}{2\ell+1}\sum_{m=-\ell}^\ell \langle n_{\ell m,i} n_{\ell m,j}^*\rangle=\delta_{ij}\sigma^2_{\ell,i} \,,
\end{equation}
where $\sigma^2_{\ell,i}$ is the mean noise power spectrum of the channel $i$.

Defining $\boldsymbol{\Omega}_\ell=\hat{\boldsymbol N}_\ell-\bar{\boldsymbol N}_\ell$, the cosmic variance of the Gaussian white noise is given by:
\begin{equation}
    \begin{aligned}
    \langle\Omega_{\ell,ij}\Omega_{\ell,i'j'}\rangle
    &=
    \frac{1}{(2\ell+1)^2}\sum_{mm'}
    (\langle n_{\ell m,i} n_{\ell m,j}^* n_{\ell m',i'} n_{\ell m',j'}^*\rangle
    -\langle n_{\ell m,i} n_{\ell m,j}^*\rangle\langle n_{\ell m',i'} n_{\ell m',j'}^*\rangle)
    \\&=
    \frac{\sigma^2_{\ell,i}\sigma^2_{\ell,j}}{2\ell+1}
    (\delta_{ii'}\delta_{jj'}+\delta_{ij'}\delta_{ji'})\,.
    \end{aligned}
\end{equation}
Defining $\boldsymbol{\Delta}_\ell=\hat{\boldsymbol C}_\ell-\bar{\boldsymbol C}_\ell$, from Eq.\eqref{eq:c^hat} and Eq.\eqref{eq:c^bar}, we can write
\begin{equation}
    \boldsymbol{\Delta}_\ell=(\hat{C}^s_{\ell}-\bar{C}^s_{\ell})\boldsymbol a\boldsymbol a^T+\boldsymbol{\Omega}_{\ell}+(\hat{\boldsymbol F}_\ell-\bar{\boldsymbol F}_\ell)+\boldsymbol{\Lambda}_\ell \,.
    \label{eq:delta_l}
\end{equation}

When we consider the correlation between $\boldsymbol{\Delta}_\ell$ and $\boldsymbol{\Omega}_\ell$, only the second term of Eq.\eqref{eq:delta_l} gives rise to the correlation. The first and the third correlation term vanish since the noise and the CMB or foregrounds are uncorrelated, while the last correlation term vanishes since it contains the third power of $\boldsymbol n_{\ell m}$ and the first power of $s_{\ell m}$. Therefore, we have
\begin{equation}
    \langle\Delta_{\ell,ij}\Omega_{\ell,i'j'}\rangle
    =
    \langle\Omega_{\ell,ij}\Omega_{\ell,i'j'}\rangle
    =
    \frac{\sigma^2_{\ell,i}\sigma^2_{\ell,j}}{2\ell+1}
    (\delta_{ii'}\delta_{jj'}+\delta_{ij'}\delta_{ji'})\,.
\end{equation}

Inserting the ILC weights to the noise bias error:
\begin{equation}
    \begin{aligned}
    \langle{N}^{\rm res.}_\ell-\bar{N}^{\rm res.}_\ell\rangle
    &=
    \langle
    \frac{\boldsymbol a^T\hat{\boldsymbol C}_{\ell}^{-1}}
    {\boldsymbol a^T\hat{\boldsymbol C}_{\ell}^{-1}\boldsymbol a}
    (\hat{\boldsymbol N}_\ell-\bar{\boldsymbol N}_\ell)
    \frac{\hat{\boldsymbol C}_{\ell}^{-1}\boldsymbol a}
    {\boldsymbol a^T\hat{\boldsymbol C}_{\ell}^{-1}\boldsymbol a}
    \rangle
    \\&=
    \langle
    \frac{\boldsymbol a^T\hat{\boldsymbol C}_{\ell}^{-1}\boldsymbol{\Omega}_\ell\hat{\boldsymbol C}_{\ell}^{-1}\boldsymbol a}
    {(\boldsymbol a^T\hat{\boldsymbol C}_{\ell}^{-1}\boldsymbol a)^2}
    \rangle\,.
    \end{aligned}
\end{equation}
Since $\boldsymbol{\Delta}_\ell$ is a small perturbation to $\hat{\boldsymbol C}_\ell$, we use the first order expansion:
\begin{equation}
    \hat{\boldsymbol C}_\ell^{-1}=[\bar{\boldsymbol C}_\ell+\boldsymbol{\Delta}_\ell]^{-1}\approx\bar{\boldsymbol C}_\ell^{-1}-\bar{\boldsymbol C}_\ell^{-1}\boldsymbol{\Delta}_\ell\bar{\boldsymbol C}_\ell^{-1}\,.
    \label{eq:Cl_expansion}
\end{equation}
Writing (Eq.(A.11) of \cite{delabrouilleFullSkyLow2009}):
\begin{equation}
    \frac{1}{\boldsymbol a^T[\bar{\boldsymbol C}_\ell^{-1}-\bar{\boldsymbol C}_\ell^{-1}\boldsymbol{\Delta}_\ell\bar{\boldsymbol C}_\ell^{-1}]\boldsymbol a}
    =
    \frac{1}{\boldsymbol a^T\bar{\boldsymbol C}_\ell^{-1}\boldsymbol a}
    \frac{1}{1-\epsilon}
    \approx
    \frac{1}{\boldsymbol a^T\bar{\boldsymbol C}_\ell^{-1}\boldsymbol a}(1+\epsilon) \,,
\end{equation}
where
\begin{equation}
    \epsilon=\frac{\boldsymbol a^T\bar{\boldsymbol C}_\ell^{-1}\boldsymbol{\Delta}_\ell\bar{\boldsymbol C}_\ell^{-1}\boldsymbol a}{\boldsymbol a^T\bar{\boldsymbol C}_\ell^{-1}\boldsymbol a}\,.
\end{equation}
The terms without $\boldsymbol{\Delta}_\ell$ vanish since $\langle\boldsymbol{\Omega}_\ell\rangle=0$. Keeping up to the first order terms of $\boldsymbol{\Delta}_\ell$:
\begin{equation}
    \begin{aligned}
    \langle{N}^{\rm res.}_\ell-\bar{N}^{\rm res.}_\ell\rangle
    \approx&
    \frac{2}{(\boldsymbol a^T\bar{\boldsymbol C}_{\ell}^{-1}\boldsymbol a)^3}
    \langle
    (\boldsymbol a^T\bar{\boldsymbol C}_\ell^{-1}\boldsymbol{\Omega}_\ell\bar{\boldsymbol C}_\ell^{-1}\boldsymbol a)
    (\boldsymbol a^T\bar{\boldsymbol C}_\ell^{-1}\boldsymbol{\Delta}_\ell\bar{\boldsymbol C}_\ell^{-1}\boldsymbol a)
    \rangle
    \\&-
    \frac{2}{(\boldsymbol a^T\bar{\boldsymbol C}_{\ell}^{-1}\boldsymbol a)^2}
    \langle
    (\boldsymbol a^T\bar{\boldsymbol C}_\ell^{-1}\boldsymbol{\Delta}_\ell\bar{\boldsymbol C}_\ell^{-1}\boldsymbol{\Omega}_\ell\bar{\boldsymbol C}_\ell^{-1}\boldsymbol a)
    \rangle\,.
    \label{eq:full-form}
    \end{aligned}
\end{equation}
For the first term of Eq.\eqref{eq:full-form}
\begin{equation}
    \begin{aligned}
    &\langle
    (\boldsymbol a^T\bar{\boldsymbol C}_\ell^{-1}\boldsymbol{\Omega}_\ell\bar{\boldsymbol C}_\ell^{-1}\boldsymbol a)
    (\boldsymbol a^T\bar{\boldsymbol C}_\ell^{-1}\boldsymbol{\Delta}_\ell\bar{\boldsymbol C}_\ell^{-1}\boldsymbol a)
    \rangle
    \\=&
    \langle
    [\sum_{ijkp}(\bar{\boldsymbol C}_\ell^{-1})_{ij}\Omega_{\ell,jk}(\bar{\boldsymbol C}_\ell^{-1})_{kp}]
    [\sum_{i'j'k'p'}(\bar{\boldsymbol C}_\ell^{-1})_{i'j'}\Delta_{\ell,j'k'}(\bar{\boldsymbol C}_\ell^{-1})_{k'p'}]
    \rangle
    \\=&
    \sum_{ijkp}(\bar{\boldsymbol C}_\ell^{-1})_{ij}(\bar{\boldsymbol C}_\ell^{-1})_{kp}\sum_{i'j'k'p'}(\bar{\boldsymbol C}_\ell^{-1})_{i'j'}(\bar{\boldsymbol C}_\ell^{-1})_{k'p'}
    \langle\Omega_{\ell,jk}\Delta_{\ell,j'k'}\rangle
    \\=&
    \frac{2}{2\ell+1}[\sum_{iji'}(\bar{\boldsymbol C}_\ell^{-1})_{ij}\sigma^{2}_{\ell,j}(\bar{\boldsymbol C}_\ell^{-1})_{ji'}][\sum_{pkp'}(\bar{\boldsymbol C}_\ell^{-1})_{pk}\sigma^{2}_{\ell,k}(\bar{\boldsymbol C}_\ell^{-1})_{kp'}]
    \\=&
    \frac{2}{2\ell+1}(\boldsymbol a^T\bar{\boldsymbol C}_{\ell}^{-1}\bar{\boldsymbol N}_{\ell}\bar{\boldsymbol C}_{\ell}^{-1}\boldsymbol a)^2
    \,.
    \end{aligned}
    \label{eq:1st-term}
\end{equation}
Here we use the symmetry of $\bar{\boldsymbol C}_\ell$. 
For the second term of Eq.\eqref{eq:full-form}
\begin{equation}
    \begin{aligned}
    &\langle
    \boldsymbol a^T\bar{\boldsymbol C}_\ell^{-1}\boldsymbol{\Delta}_\ell\bar{\boldsymbol C}_\ell^{-1}\boldsymbol{\Omega}_\ell\bar{\boldsymbol C}_\ell^{-1}\boldsymbol a
    \rangle
    \\=&
    \sum_{ijkpqr}(\bar{\boldsymbol C}_\ell^{-1})_{ij}(\bar{\boldsymbol C}_\ell^{-1})_{kp}(\bar{\boldsymbol C}_\ell^{-1})_{qr}\langle\Delta_{\ell,jk}\Omega_{\ell,qr}\rangle
    \\=&
    \frac{1}{2\ell+1}\sum_{ijkr}(\bar{\boldsymbol C}_\ell^{-1})_{ij}\sigma^{2}_{\ell,j}\sigma^{2}_{\ell,k}[(\bar{\boldsymbol C}_\ell^{-1})_{jk}(\bar{\boldsymbol C}_\ell^{-1})_{kr}+(\bar{\boldsymbol C}_\ell^{-1})_{kk}(\bar{\boldsymbol C}_\ell^{-1})_{jr}]
    \\=&
    \frac{1}{2\ell+1}[\boldsymbol a^T\bar{\boldsymbol C}_\ell^{-1}\bar{\boldsymbol N}_\ell\bar{\boldsymbol C}_\ell^{-1}\bar{\boldsymbol N}_\ell\bar{\boldsymbol C}_\ell^{-1}\boldsymbol a+\boldsymbol a^T\bar{\boldsymbol C}_\ell^{-1}\bar{\boldsymbol N}_\ell\bar{\boldsymbol C}_\ell^{-1}\boldsymbol a\ {\rm Tr}(\bar{\boldsymbol N}_\ell\bar{\boldsymbol C}_\ell^{-1})]
    \,.
    \end{aligned}
    \label{eq:2nd-term}
\end{equation}
Ignoring the foreground term we have $\bar{\boldsymbol N}_\ell=\bar{\boldsymbol C}_\ell-\bar{C}^s_{\ell}\boldsymbol a\boldsymbol a^T$. Using
\begin{equation}
    \begin{aligned}
    \boldsymbol a^T\bar{\boldsymbol C}_\ell^{-1}\bar{\boldsymbol N}_\ell\bar{\boldsymbol C}_\ell^{-1}\bar{\boldsymbol N}_\ell\bar{\boldsymbol C}_\ell^{-1}\boldsymbol a
    &\approx
    \boldsymbol a^T\bar{\boldsymbol C}_\ell^{-1}\bar{\boldsymbol N}_\ell\bar{\boldsymbol C}_\ell^{-1}(\bar{\boldsymbol C}_\ell-\bar{C}^s_{\ell}\boldsymbol a\boldsymbol a^T)\bar{\boldsymbol C}_\ell^{-1}\boldsymbol a
    \\&=
    (\boldsymbol a^T\bar{\boldsymbol C}_\ell^{-1}\bar{\boldsymbol N}_\ell\bar{\boldsymbol C}_\ell^{-1}\boldsymbol a)[1-\bar{C}^s_{\ell}(\boldsymbol a^T\bar{\boldsymbol C}_\ell^{-1}\boldsymbol a)]
    \,,
    \end{aligned}
\end{equation}
and
\begin{equation}
    \begin{aligned}
    {\rm Tr}(\bar{\boldsymbol N}_\ell\bar{\boldsymbol C}_\ell^{-1})
    &\approx
    {\rm Tr}[(\bar{\boldsymbol C}_\ell-\bar{C}^s_{\ell}\boldsymbol a\boldsymbol a^T)\bar{\boldsymbol C}_\ell^{-1}]
    \\&=
    {\rm Tr}({\boldsymbol I})-\bar{C}^s_{\ell}{\rm Tr}(\boldsymbol a\boldsymbol a^T\bar{\boldsymbol C}_\ell^{-1})
    \\&=
    n_\nu-\bar{C}^s_{\ell}(\boldsymbol a^T\bar{\boldsymbol C}_\ell^{-1}\boldsymbol a)
    \,,
    \end{aligned}
    \label{eq:trace}
\end{equation}
Eq.\eqref{eq:2nd-term} turns out to be
\begin{equation}
    \begin{aligned}
    &\langle
    \boldsymbol a^T\bar{\boldsymbol C}_\ell^{-1}\boldsymbol{\Delta}_\ell\bar{\boldsymbol C}_\ell^{-1}\boldsymbol{\Omega}_\ell\bar{\boldsymbol C}_\ell^{-1}\boldsymbol a
    \rangle
    \\=&
    \frac{1}{2\ell+1}(\boldsymbol a^T\bar{\boldsymbol C}_\ell^{-1}\bar{\boldsymbol N}_\ell\bar{\boldsymbol C}_\ell^{-1}\boldsymbol a)[n_\nu+1-2\bar{C}^s_{\ell}(\boldsymbol a^T\bar{\boldsymbol C}_\ell^{-1}\boldsymbol a)]
    \,.
    \end{aligned}
    \label{eq:2nd-term-final}
\end{equation}
Bring Eq.\eqref{eq:1st-term} and Eq.\eqref{eq:2nd-term-final} to Eq.\eqref{eq:full-form}:
\begin{equation}
    \begin{aligned}
    \langle{N}^{\rm res.}_\ell-\bar{N}^{\rm res.}_\ell\rangle
    \approx&
    \frac{4}{2\ell+1}\frac{(\boldsymbol a^T\bar{\boldsymbol C}_\ell^{-1}\bar{\boldsymbol N}_\ell\bar{\boldsymbol C}_\ell^{-1}\boldsymbol a)^2}{(\boldsymbol a^T\bar{\boldsymbol C}_{\ell}^{-1}\boldsymbol a)^3}
    -
    \frac{2(n_\nu+1)}{2\ell+1}\frac{(\boldsymbol a^T\bar{\boldsymbol C}_\ell^{-1}\bar{\boldsymbol N}_\ell\bar{\boldsymbol C}_\ell^{-1}\boldsymbol a)}{(\boldsymbol a^T\bar{\boldsymbol C}_{\ell}^{-1}\boldsymbol a)^2}
    \\&+
    \frac{4}{2\ell+1}\bar{C}^s_{\ell}\frac{\boldsymbol a^T\bar{\boldsymbol C}_\ell^{-1}\bar{\boldsymbol N}_\ell\bar{\boldsymbol C}_\ell^{-1}\boldsymbol a}{\boldsymbol a^T\bar{\boldsymbol C}_{\ell}^{-1}\boldsymbol a}
    \,. 
    \end{aligned}
\end{equation}
Given the mean residual noise $\bar{N}^{\rm res.}_\ell\approx\frac{\boldsymbol a^T\bar{\boldsymbol C}_{\ell}^{-1}\bar{\boldsymbol N}_{\ell}\bar{\boldsymbol C}_{\ell}^{-1}\boldsymbol a}{(\boldsymbol a^T\bar{\boldsymbol C}_{\ell}^{-1}\boldsymbol a)^2}$ and the mean power spectrum of the ILC map $\bar{C}^{\rm ILC}_\ell=\langle\hat{\boldsymbol w}^T\hat{\boldsymbol C}_\ell\hat{\boldsymbol w}\rangle\approx\frac{1}{\boldsymbol a^T\bar{\boldsymbol C}_{\ell}^{-1}\boldsymbol a}$, replacing $\bar{C}^s_{\ell}$ with $(\bar{C}^{\rm ILC}_\ell-\bar{N}^{\rm res.}_\ell)$, we get the final form of the noise bias error
\begin{equation}
    \langle{N}^{\rm res.}_\ell-\bar{N}^{\rm res.}_\ell\rangle
    \approx
    -\frac{2(n_\nu-1)}{2\ell+1}\bar{N}^{\rm res.}_\ell
    \,. 
\end{equation}
Considering the binning and partial-sky effects, we get
\begin{equation}
    \langle{N}^{\rm res.}_\ell-\bar{N}^{\rm res.}_\ell\rangle
    \approx
    -\frac{2(n_\nu-1)}{(2\ell+1)\Delta\ell_b f_{\rm sky}}\bar{N}^{\rm res.}_\ell\,,
\end{equation}
where $\Delta\ell_b$ is the binning size of the empirical covariance matrix and $f_{\rm sky}$ the sky fraction. The general form in all domains (harmonic, pixel and needlet) is
\begin{equation}
    \langle{N}^{\rm res.}_\ell-\bar{N}^{\rm res.}_\ell\rangle
    \approx
    -\frac{2(n_\nu-1)}{N_k f_{\rm sky}}\bar{N}^{\rm res.}_\ell\,.
\end{equation}

\subsection{The noise bias error in cILC}
The data is modelled as $\boldsymbol d_{\ell m}=\boldsymbol A \boldsymbol s_{\ell m}+\boldsymbol n_{\ell m}$ where $\boldsymbol A$ is the $n_\nu\times n_c$ mixing matrix and $\boldsymbol s_{\ell m}$ is the 3-dim vector of three sky components, the CMB, the thermal dust and the synchrotron. The cILC weights are given by:
\begin{equation}
    \hat{\boldsymbol w}^T=\boldsymbol{e}^T(\boldsymbol{A}^T\hat{\boldsymbol C}_{\ell}^{-1}\boldsymbol{A})^{-1}\boldsymbol{A}^T\hat{\boldsymbol C}_{\ell}^{-1}\,,
\end{equation}
where $\boldsymbol{e}=[1,0,0]^T$. The noise bias error is
\begin{equation}
    \langle{N}^{\rm res.}_\ell-\bar{N}^{\rm res.}_\ell\rangle=\langle\boldsymbol{e}^T(\boldsymbol{A}^T\hat{\boldsymbol C}_{\ell}^{-1}\boldsymbol{A})^{-1}\boldsymbol{A}^T\hat{\boldsymbol C}_{\ell}^{-1}\boldsymbol{\Omega}_\ell\hat{\boldsymbol C}_{\ell}^{-1}\boldsymbol{A}(\boldsymbol{A}^T\hat{\boldsymbol C}_{\ell}^{-1}\boldsymbol{A})^{-1}\boldsymbol{e}\rangle\,.
\end{equation}
Using Eq.\eqref{eq:Cl_expansion}, 
\begin{equation}
    \begin{aligned}
    (\boldsymbol{A}^T\hat{\boldsymbol C}_{\ell}^{-1}\boldsymbol{A})^{-1}
    &\approx
    (\boldsymbol{A}^T\bar{\boldsymbol C}_{\ell}^{-1}\boldsymbol{A}-\boldsymbol{A}^T\bar{\boldsymbol C}_{\ell}^{-1}\boldsymbol{\Delta}_{\ell}\bar{\boldsymbol C}_{\ell}^{-1}\boldsymbol{A})^{-1}
    \\&=
    (\boldsymbol{A}^T\bar{\boldsymbol C}_{\ell}^{-1}\boldsymbol{A})^{-1}(\boldsymbol{I}-\boldsymbol{\epsilon})^{-1}
    \\&\approx
    (\boldsymbol{A}^T\bar{\boldsymbol C}_{\ell}^{-1}\boldsymbol{A})^{-1}(\boldsymbol{I}+\boldsymbol{\epsilon})\,,
    \end{aligned}
\end{equation}
where
\begin{equation}
    \boldsymbol{\epsilon}=
    \boldsymbol{A}^T\bar{\boldsymbol C}_{\ell}^{-1}\boldsymbol{\Delta}_{\ell}\bar{\boldsymbol C}_{\ell}^{-1}\boldsymbol{A}(\boldsymbol{A}^T\bar{\boldsymbol C}_{\ell}^{-1}\boldsymbol{A})^{-1}\,.
\end{equation}
Keeping up to the first order terms of $\boldsymbol{\Delta}_\ell$, the noise bias error includes two terms, of which the first term:
\begin{equation}
    \begin{aligned}
    &2\langle\boldsymbol{e}^T(\boldsymbol{A}^T\bar{\boldsymbol C}_{\ell}^{-1}\boldsymbol{A})^{-1}\boldsymbol{\epsilon}\boldsymbol{A}^T\bar{\boldsymbol C}_{\ell}^{-1}\boldsymbol{\Omega}_\ell\bar{\boldsymbol C}_{\ell}^{-1}\boldsymbol{A}(\boldsymbol{A}^T\bar{\boldsymbol C}_{\ell}^{-1}\boldsymbol{A})^{-1}\boldsymbol{e}\rangle
    \\=&
    2\sum_{ijkp}(\boldsymbol{e}^T(\boldsymbol{A}^T\bar{\boldsymbol C}_{\ell}^{-1}\boldsymbol{A})^{-1}\boldsymbol{A}^T\bar{\boldsymbol C}_{\ell}^{-1})_{i}\langle{\Delta}_{\ell,ij}{\Omega}_{\ell,kp}\rangle
    \\&
    (\bar{\boldsymbol C}_{\ell}^{-1}\boldsymbol{A}(\boldsymbol{A}^T\bar{\boldsymbol C}_{\ell}^{-1}\boldsymbol{A})^{-1}\boldsymbol{A}^T\bar{\boldsymbol C}_{\ell}^{-1})_{jk}(\bar{\boldsymbol C}_{\ell}^{-1}\boldsymbol{A}(\boldsymbol{A}^T\bar{\boldsymbol C}_{\ell}^{-1}\boldsymbol{A})^{-1}\boldsymbol{e})_{p}
    \\=&
    \frac{2}{2\ell+1}(
    \boldsymbol{e}^T(\boldsymbol{A}^T\bar{\boldsymbol C}_{\ell}^{-1}\boldsymbol{A})^{-1}\boldsymbol{A}^T\bar{\boldsymbol C}_{\ell}^{-1}\bar{\boldsymbol N}_{\ell}\bar{\boldsymbol C}_{\ell}^{-1}\boldsymbol{A}(\boldsymbol{A}^T\bar{\boldsymbol C}_{\ell}^{-1}\boldsymbol{A})^{-1}\boldsymbol{A}^T\bar{\boldsymbol C}_{\ell}^{-1}\bar{\boldsymbol N}_{\ell}\bar{\boldsymbol C}_{\ell}^{-1}\boldsymbol{A}(\boldsymbol{A}^T\bar{\boldsymbol C}_{\ell}^{-1}\boldsymbol{A})^{-1}\boldsymbol{e}
    \\&+
    \boldsymbol{e}^T(\boldsymbol{A}^T\bar{\boldsymbol C}_{\ell}^{-1}\boldsymbol{A})^{-1}\boldsymbol{A}^T\bar{\boldsymbol C}_{\ell}^{-1}\bar{\boldsymbol N}_{\ell}\bar{\boldsymbol C}_{\ell}^{-1}\boldsymbol{A}(\boldsymbol{A}^T\bar{\boldsymbol C}_{\ell}^{-1}\boldsymbol{A})^{-1}\boldsymbol{e}{\rm Tr}[\bar{\boldsymbol N}_{\ell}\bar{\boldsymbol C}_{\ell}^{-1}\boldsymbol{A}(\boldsymbol{A}^T\bar{\boldsymbol C}_{\ell}^{-1}\boldsymbol{A})^{-1}\boldsymbol{A}^T\bar{\boldsymbol C}_{\ell}^{-1}])\,.
    \end{aligned}
    \label{eq:1st-term-cilc}
\end{equation}
Inserting $\bar{\boldsymbol N}_{\ell}=\bar{\boldsymbol C}_{\ell}-\boldsymbol{A}\bar{\boldsymbol S}_{\ell}\boldsymbol{A}^T$ where $\bar{\boldsymbol S}_{\ell}=\frac{1}{2\ell+1}\sum_{m=-\ell}^{\ell} \langle\boldsymbol s_{\ell m} \boldsymbol s_{\ell m}^\dagger\rangle$, the first term of Eq.\eqref{eq:1st-term-cilc} becomes
\begin{equation}
    \frac{2}{2\ell+1}[\boldsymbol{e}^T(\boldsymbol{A}^T\bar{\boldsymbol C}_{\ell}^{-1}\boldsymbol{A})^{-1}\boldsymbol{e}-2\boldsymbol{e}^T\bar{\boldsymbol S}_{\ell}\boldsymbol{e}+\boldsymbol{e}^T\bar{\boldsymbol S}_{\ell}(\boldsymbol{A}^T\bar{\boldsymbol C}_{\ell}^{-1}\boldsymbol{A})\bar{\boldsymbol S}_{\ell}\boldsymbol{e}]\,,
\end{equation}
and the trace term of Eq.\eqref{eq:1st-term-cilc}
\begin{equation}
    {\rm Tr}[\boldsymbol{A}(\boldsymbol{A}^T\bar{\boldsymbol C}_{\ell}^{-1}\boldsymbol{A})^{-1}\boldsymbol{A}^T\bar{\boldsymbol C}_{\ell}^{-1}-\boldsymbol{A}\bar{\boldsymbol S}_{\ell}\boldsymbol{A}^T\bar{\boldsymbol C}_{\ell}^{-1}]
    =n_c-{\rm Tr}[\boldsymbol{A}\bar{\boldsymbol S}_{\ell}\boldsymbol{A}^T\bar{\boldsymbol C}_{\ell}^{-1}]\,.
\end{equation}
The second term of the noise bias error:
\begin{equation}
    \begin{aligned}
    &-2\langle\boldsymbol{e}^T(\boldsymbol{A}^T\bar{\boldsymbol C}_{\ell}^{-1}\boldsymbol{A})^{-1}\boldsymbol{A}^T\bar{\boldsymbol C}_{\ell}^{-1}\boldsymbol {\Delta}_{\ell}\bar{\boldsymbol C}_{\ell}^{-1}\boldsymbol{\Omega}_{\ell}\bar{\boldsymbol C}_{\ell}^{-1}\boldsymbol{A}(\boldsymbol{A}^T\bar{\boldsymbol C}_{\ell}^{-1}\boldsymbol{A})^{-1}\boldsymbol{e}\rangle
    \\=&
    -2\sum_{ijkp}(\boldsymbol{e}^T(\boldsymbol{A}^T\bar{\boldsymbol C}_{\ell}^{-1}\boldsymbol{A})^{-1}\boldsymbol{A}^T\bar{\boldsymbol C}_{\ell}^{-1})_i\langle{\Delta}_{\ell,ij}{\Omega}_{\ell,kp}\rangle(\bar{\boldsymbol C}_{\ell}^{-1})_{jk}
    (\bar{\boldsymbol C}_{\ell}^{-1}\boldsymbol{A}(\boldsymbol{A}^T\bar{\boldsymbol C}_{\ell}^{-1}\boldsymbol{A})^{-1}\boldsymbol{e})_p
    \\=&
    -\frac{2}{2\ell+1}(\boldsymbol{e}^T(\boldsymbol{A}^T\bar{\boldsymbol C}_{\ell}^{-1}\boldsymbol{A})^{-1}\boldsymbol{A}^T\bar{\boldsymbol C}_{\ell}^{-1}\bar{\boldsymbol N}_{\ell}\bar{\boldsymbol C}_{\ell}^{-1}\bar{\boldsymbol N}_{\ell}\bar{\boldsymbol C}_{\ell}^{-1}\boldsymbol{A}(\boldsymbol{A}^T\bar{\boldsymbol C}_{\ell}^{-1}\boldsymbol{A})^{-1}\boldsymbol{e}
    \\&+
    \boldsymbol{e}^T(\boldsymbol{A}^T\bar{\boldsymbol C}_{\ell}^{-1}\boldsymbol{A})^{-1}\boldsymbol{A}^T\bar{\boldsymbol C}_{\ell}^{-1}\bar{\boldsymbol N}_{\ell}\bar{\boldsymbol C}_{\ell}^{-1}\boldsymbol{A}(\boldsymbol{A}^T\bar{\boldsymbol C}_{\ell}^{-1}\boldsymbol{A})^{-1}\boldsymbol{e}{\rm Tr}[\bar{\boldsymbol N}_{\ell}\bar{\boldsymbol C}_{\ell}^{-1}])\,.
    \end{aligned}
    \label{eq:2nd-term-cilc}
\end{equation}
Inserting $\bar{\boldsymbol N}_{\ell}=\bar{\boldsymbol C}_{\ell}-\boldsymbol{A}\bar{\boldsymbol S}_{\ell}\boldsymbol{A}^T$, the first term of Eq.\eqref{eq:2nd-term-cilc} becomes
\begin{equation}
    -\frac{2}{2\ell+1}[\boldsymbol{e}^T(\boldsymbol{A}^T\bar{\boldsymbol C}_{\ell}^{-1}\boldsymbol{A})^{-1}\boldsymbol{e}-2\boldsymbol{e}^T\bar{\boldsymbol S}_{\ell}\boldsymbol{e}+\boldsymbol{e}^T\bar{\boldsymbol S}_{\ell}(\boldsymbol{A}^T\bar{\boldsymbol C}_{\ell}^{-1}\boldsymbol{A})\bar{\boldsymbol S}_{\ell}\boldsymbol{e}]\,,
\end{equation}
and the trace term of Eq.\eqref{eq:2nd-term-cilc}
\begin{equation}
    {\rm Tr}[\boldsymbol{I}-\boldsymbol{A}\bar{\boldsymbol S}_{\ell}\boldsymbol{A}^T\bar{\boldsymbol C}_{\ell}^{-1}]
    =n_\nu-{\rm Tr}[\boldsymbol{A}\bar{\boldsymbol S}_{\ell}\boldsymbol{A}^T\bar{\boldsymbol C}_{\ell}^{-1}]\,.
\end{equation}
Summing all the terms above, the noise bias error
\begin{equation}
    \langle{N}^{\rm res.}_\ell-\bar{N}^{\rm res.}_\ell\rangle\approx\frac{2(n_c-n_\nu)}{2\ell+1}\boldsymbol{e}^T(\boldsymbol{A}^T\bar{\boldsymbol C}_{\ell}^{-1}\boldsymbol{A})^{-1}\boldsymbol{A}^T\bar{\boldsymbol C}_{\ell}^{-1}\bar{\boldsymbol N}_{\ell}\bar{\boldsymbol C}_{\ell}^{-1}\boldsymbol{A}(\boldsymbol{A}^T\bar{\boldsymbol C}_{\ell}^{-1}\boldsymbol{A})^{-1}\boldsymbol{e}\,.
\end{equation}
Given the mean residual noise $\bar{N}^{\rm res.}_\ell\approx\boldsymbol{e}^T(\boldsymbol{A}^T\bar{\boldsymbol C}_{\ell}^{-1}\boldsymbol{A})^{-1}\boldsymbol{A}^T\bar{\boldsymbol C}_{\ell}^{-1}\bar{\boldsymbol N}_{\ell}\bar{\boldsymbol C}_{\ell}^{-1}\boldsymbol{A}(\boldsymbol{A}^T\bar{\boldsymbol C}_{\ell}^{-1}\boldsymbol{A})^{-1}\boldsymbol{e}$, the general form of the noise bias error for cILC is
\begin{equation}
    \langle{N}^{\rm res.}_\ell-\bar{N}^{\rm res.}_\ell\rangle\approx\frac{2(n_c-n_\nu)}{N_kf_{\rm sky}}\bar{N}^{\rm res.}_\ell\,.
\end{equation}

A comparison between the theoretical amplitude and the actual error from simulations has been plotted in the lower panel of Figure~\ref{fig:nbe}, indicating that the above approximation is sensible.

\section{\revise{Verifying NBE of Jackknife}}
\label{sec:ver_nbe}

\begin{figure}[tbp]
    \centering
    \begin{subfigure}[b]{0.8\textwidth}
        \centering
        \includegraphics[width=\textwidth]{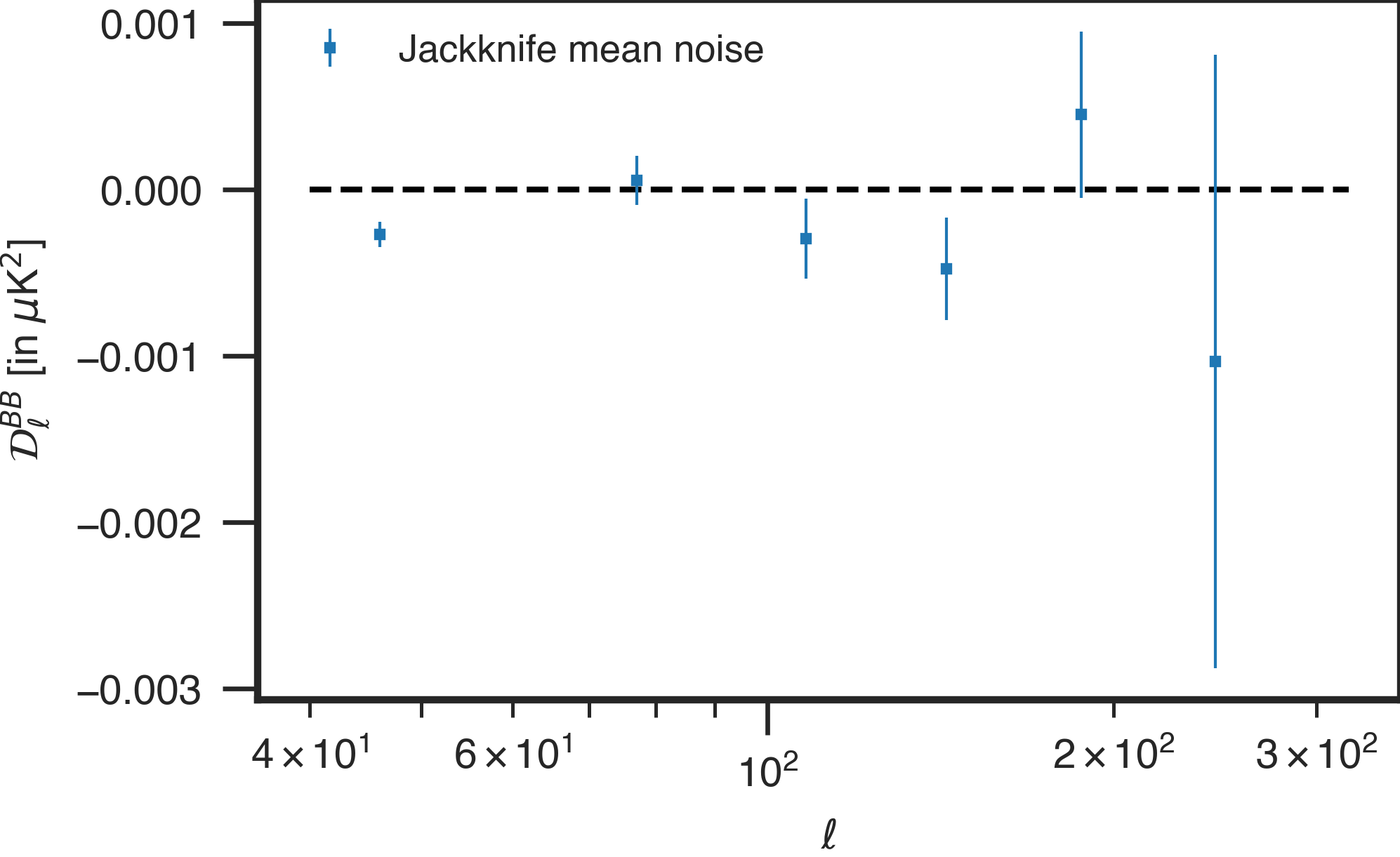}
    \end{subfigure}
    \caption{The mean residual noise cross spectrum for the Jackknife test. The plotted error bars are $\sigma(\bar{N}^{\rm res.}_\ell)=\sigma({N}^{\rm res.}_\ell)/10$, denoting the uncertainty of the noise bias averaged over 100 noise simulations.}
    \label{fig:nbe-jack}
\end{figure}

\revise{As discussed in section~\ref{sec:nbe}, we are about to verify that the Jackknife method would produce no noise bias error. First we add the same CMB and baseline foregrounds to different noise realizations ($\sqrt{2}$ of the original level) to generate the mock data splits. In the Jackknife pipeline, we apply the same cILC weights obtained from the combined data averaging two data splits onto two splits respectively, and finally compute the cross power spectrum between two cILC-cleaned split maps. The averaged residual noise power spectrum and its error bar over 100 simulations are plotted in Figure~\ref{fig:nbe-jack}. Indeed, the mean noise power spectrum for Jackknife is consistent with zero at all scales (except for the largest one, where the multipole modes are not enough to manifest the expectation).}

\acknowledgments
We appreciate the helpful discussions in the AliCPT science
team.
This work is supported by the National Key R\&D Program of China Grant No. 2021YFC2203102 and 2022YFC2204602, Strategic Priority Research Program of the Chinese
Academy of Science Grant No. XDB0550300,
NSFC No. 12325301 and 12273035, the Fundamental Research Funds for the Central Universities under Grant No. WK3440000004, the China Manned Space Project with No.CMS-CSST-2021-B01, and the 111 Project for "Observational and Theoretical Research on Dark Matter and Dark Energy" (B23042). L.S. is supported by the National Key R\&D Program of China (2020YFC2201600) and NSFC grant 12150610459.


    
    


\bibliography{Bib}
\bibliographystyle{JHEP}

\end{document}